\documentclass[10pt,a4paper]{article}
\usepackage{jheppub_kim}
\usepackage{pdflscape}
\usepackage{amsmath}
\usepackage{amssymb}
\usepackage{dcolumn}
\usepackage{bm}
\usepackage{color}
\usepackage{epsfig}
\usepackage{amsfonts}
\usepackage{graphicx}
\usepackage{subfigure}
\usepackage{dcolumn}

\begin{document}
\title{A Time Dependent Spacetime in $f(R,T)$ Gravity: Gravitational Collapse}

\author[a]{Prabir Rudra}

\affiliation[a] {Department of Mathematics, Asutosh College,
Kolkata-700 026, India}

\emailAdd{prudra.math@gmail.com, rudra@associates.iucaa.in}

\abstract{In this note a time dependent spacetime is explored in
the background of $f(R,T)$ gravity via the gravitational collapse
of a massive star. The star is modelled by the Vaidya spacetime
which is time dependent in nature. The coupling of matter with
curvature is the key feature of $f(R,T)$ theory and here we have
investigated its effects on a collapsing scenario. Two different
types of models, one involving minimal and the other involving
non-minimal coupling between matter and curvature are considered
for our study. Power law and exponential functionalities are
considered as examples to check the outcome of the gravitational
collapse. Our prime objective is to explore the nature of
singularities (black hole or naked singularity) that form as an
end state of the collapse. Existence of outgoing radial null
geodesics from the central singularity was probed and such
existence implied the formation of naked singularities thus
defying the cosmic censorship hypothesis. The absence of such
outgoing null geodesics would imply the formation of an event
horizon and the singularity formed becomes a black hole.
Conditions under which such possibilities occur are derived for
all the models and sub-models. Gravitational strength of the
singularity is also investigated and the conditions under which we
can get a strong or a weak singularity is derived. The results
obtained are very interesting and may be attributed to the
coupling between curvature and matter. It is seen that for
non-minimal coupling there is a possibility of a globally naked
singularity, whereas for a minimal coupling scenario local
nakedness is the only option. It is also found that the
singularity formed can be sufficiently weak in nature, which is
cosmologically desirable.}


\maketitle

\section{Introduction}
For the last two decades we have been aware of the fact that our
universe has entered into a phase of accelerated expansion
\cite{acc1, acc2}. Although this came as a total surprise to the
scientific community, extensive research have been able to put
some meaning to this observed phenomenon over the years. It is a
widely known fact that this cosmic acceleration can be explained
via two different theoretical frameworks. One is the theory of
\textit{dark energy (DE)} that recasts the matter content of the
universe to some exotic substance possessing negative pressure.
The other way is to modify Einstein's theory of gravity leading to
\textit{modified gravity theories}. The reader may refer to
Refs.\cite{mod1, mod2, mod3} for extensive reviews on modified
gravity theories and to the Ref.\cite{de1} for a detailed review
on DE.

One of the most popular way to modify Einstein's gravity is by
replacing the Ricci scalar $R$ in the gravity Lagrangian of the
Einstein-Hilbert action of general relativity (GR) by an analytic
function of $R$, i.e. $f(R)$, which gives rise to $f(R)$ gravity
theory. Extensive reviews on $f(R)$ gravity can be found in
Refs.\cite{fr1, fr2}. In Ref.\cite{frlm} the authors proposed an
even more generic class of models by considering the gravitational
lagrangian as an analytic function of Ricci scalar $R$ and matter
Lagrangian $L_{m}$, paving the path for $f(R, L_{m})$ theories.
Further developments in $f(R, L_{m})$ theories can be found in
Refs.\cite{frlm2, frlm3, frlm4}. In Ref.\cite{harko1} Harko et al.
proposed the $f(R,T)$ theory, where the matter Lagrangian is given
by the trace $T$ of the energy-momentum tensor $T_{\mu\nu}$. So
here the gravitational Lagrangian is an analytic function of two
scalar invariants, namely the Ricci scalar $R$ and the trace of
the energy-momentum tensor $T$. Here the contributions of $T$ will
come from the matter content of the universe. As a result it is
found that the field equations of $f(R,T)$ theory depends on a
source term, which is given by the variation of the
energy-momentum tensor with respect to the metric. This will in
turn depend on the matter Lagrangian or the nature of matter
content of the universe. So it is obvious that for different types
of matter, such as scalar fields, perfect fluid, electromagnetic
field, etc. we will get different set of field equations. From the
form of the function, it is obvious that this theory involves
coupling between matter and geometry. So by studying this theory
one can probe such coupling effects and their consequences on
various astrophysical and cosmological phenomenon. It is seen that
the covariant divergence of the energy-momentum tensor is non-zero
for this model, which leads to non-geodesic motion of the massive
test particles. This is because the coupling effects between
matter and geometry induces an extra acceleration on the
particles. Thermodynamics in $f(R,T)$ gravity was studied by
Sharif and Zubair in \cite{frt1}. Cosmological Evolution in
$f(R,T)$ theory with collisional matter was studied in
Ref.\cite{frt2}. In Ref.\cite{frt3} cosmic coincidence problem was
studied in the background of $f(R,T)$ gravity. Cosmological models
in $f(R,T)$ theories as phase space was explored in \cite{frt4}.
Dynamics of scalar perturbations in $f(R,T)$ gravity was studied
by Alvarenga et al. in \cite{frt5}. Gravastars in $f(R,T)$ gravity
was studied in Ref.\cite{frt6}. Dark matter from $f(R,T)$ gravity
was investigated by the authors in \cite{frt7}. Propagation of
polar gravitational waves in $f(R,T)$ scenario was explored in
\cite{frt8}. Dynamical behavior of the Tolman metrics in $f(R,T)$
gravity was studied by Hansraj and Banerjee in \cite{frt9}.

Gravitational collapse is a key astrophysical phenomenon that
helps us to understand various aspects of the universe such as
structure formation, properties of stars, formation of black
holes, white dwarfs, neutron stars, etc. A star undergoes a
gravitational collapse due to its own mass at the end of its life
cycle, when it has exhausted all its nuclear fuel. During its
collapse journey there are various stages at which the collapse
may stop, depending on the initial mass of the collapsing star. If
the star is massive i.e. mass$>20 M_{\odot}$ ($M_{\odot}$
represents solar mass), then the collapse does not come to a halt
at any of the intermediate stages (such as white dwarf or neutron
star), but directly proceeds to form a singularity such as a black
hole (BH). The study of gravitational collapse started with
Oppenheimer and Snyder \cite{oppen} in $1939$ when they explored
the gravitational collapse of a dust cloud modelled by a static
Schwarzschild exterior and Friedmann interior. Following this,
Tolman \cite{tolman} and Bondi \cite{bondi} studied the collapse
of spherically symmetric inhomogeneous distribution of dust.
Subsequently a lot of interest was generated in this subject and
numerous work related to this can be found in literature. Some
reviews in gravitational collapse can be found in
Refs.\cite{collrev1, collrev2}. Roger Penrose in 1969 proposed
cosmic censorship hypothesis (CCH) \cite{penrose}, where he stated
that any cosmological singularity will always be covered by an
event horizon, thus censoring the singularity from an external
observer. Such a singularity (popularly called a \textit{black
hole}) is associated with permanent loss of physical information
allowing multiple physical states to devolve into a single state.
This is known as the \textit{black hole information loss paradox}
\cite{paradox1, paradox2, paradox3, paradox4}. Over the years, in
the absence of a formidable proof of CCH, scientists started
questioning its validity. As a consequence, a search was initiated
that will culminate in the discovery of a singularity that will be
free from any event horizon. This type od singularity will not
only disprove CCH but also in the absence of information loss it
will enhance our knowledge about gravity. Such a singularity is
named as a \textit{naked singularity (NS)} \cite{ns1, ns2, ns3,
ns4, ns5, cch2, cch3, ns6, ns7, lake2, szek, ghosh1} which is
considered to be a crucial tool in the formulation of an effective
theory of quantum gravity.

The first effective relativistic line element representing the
spacetime of a realistic star was given by P. C. Vaidya \cite{v1}
in 1951. It represented the radiation for a non-static mass, thus
generalizing the static solution of Schwarzschild. Schwarzschild's
solution basically represented the spacetime around a spherically
symmetric cold dark body with a constant mass. So it is obvious
that it could never model the spacetime outside a star. This is
the problem that was addressed by Vaidya in his phenomenal paper
\cite{v1} of 1951. The solution proposed by Vaidya was termed as
Vaidya spacetime and is often referred to as the \textit{shining
or radiating Schwarzschild metric}. It should be noted that the
basic difference between the two metrics is that the constant mass
parameter in the Schwarzschild metric is replaced by a time
dependent mass parameter in the Vaidya metric, which consequently
becomes a time dependent spacetime. Notable studies in Vaidya
metric can be found in the Refs.\cite{s1, s2, s3, s4, s5, s6, s7,
s8, s9}.

Here we are interested in exploring the gravitational collapse of
a massive star modelled by the Vaidya metric in the background of
$f(R,T)$ gravity. Collapsing scenario in the presence of coupling
between matter and curvature is expected to be an interesting
proposition. Moreover the behaviour of Vaidya spacetime has never
been explored in the background of $f(R,T)$ gravity. So there is
more than enough motivation for attempting this work. We will
basically focus on the nature of the singularity formed (BH or NS)
as the end state of the collapse. We will report the conditions
under which these singularities can form in a comparative manner.
We hope to obtain interesting and new results in our collapsing
scheme in the background of curvature-matter coupling. In the next
section we will report the basic equations of Vaidya spacetime in
$f(R,T)$ gravity and find solutions for the system. In section III
we will explore the collapsing scenario of a massive star. Section
IV will deal with the strength of the singularity formed and
finally the paper will end with a detailed discussion and
conclusion in section V.

\section{Vaidya spacetime in $f(R,T)$ gravity}
The Einstein-Hilbert action for general relativity is given by,
\begin{equation}\label{actionEH}
S_{EH}=\frac{1}{2\kappa}\int R\sqrt{-g}d^{4}x
\end{equation}
where $\kappa\equiv 8\pi$, $g$ is the determinant of the metric
and $R$ is the Ricci scalar (we have considered $G$=$c=1$). We
replace the Ricci scalar, $R$ in the above action by a generalized
function of $R$ to get the action for $f(R)$ gravity \cite{fr1,
fr2},
\begin{equation}\label{actionR}
S=\frac{1}{2\kappa}\int f(R)\sqrt{-g}d^{4}x
\end{equation}
Taking the action (\ref{actionR}) and adding a matter term $S_M$,
the total action for $f(R)$ gravity takes the form,
\begin{equation}\label{actionRtotal}
S_{f(R)}=\frac{1}{2\kappa}\int f(R)\sqrt{-g}d^{4}x+\int
\mathcal{L}_{m}\sqrt{-g}d^{4}x
\end{equation}
where $\mathcal{L}_m$ is the matter Lagrangian and the second
integral on the R.H.S is $S_M$ representing the matter fields. To
obtain the action for $f(R,T)$ gravity we further modify the
action for $f(R)$ gravity by introducing the trace of the
energy-momentum tensor $T_{\mu\nu}$ in the gravity Lagrangian as
follows \cite{harko1},
\begin{equation}\label{action}
S_{f(R,T)}=\frac{1}{2\kappa}\int f(R,T)\sqrt{-g}d^{4}x+\int
\mathcal{L}_{m}\sqrt{-g}d^{4}x
\end{equation}
Here $f(R,T)$ is an arbitrary function of the Ricci scalar $R$ and
the trace $T$ of the energy-momentum tensor $T_{\mu\nu}$. The
energy-momentum tensor is defined as \cite{land1},
\begin{equation}\label{energymomentum}
T_{\mu\nu}=-\frac{2}{\sqrt{-g}}\frac{\delta
(\sqrt{-g}\mathcal{L}_{m})}{\delta g^{\mu\nu}}
\end{equation}
The trace of this tensor can be given as $T=g^{\mu\nu}T_{\mu\nu}$.
Taking variation with respect to the metric we get the field
equations for $f(R,T)$ gravity as,
\begin{equation}\label{field}
f_{R}(R,T)R_{\mu\nu}-\frac{1}{2}f(R,T)g_{\mu\nu}+\left(g_{\mu\nu}\Box-\nabla_{\mu}\nabla_{\nu}\right)f_{R}(R,T)=\kappa
T_{\mu\nu}-f_{T}(R,T)T_{\mu\nu}-f_{T}(R,T)\Theta_{\mu\nu}
\end{equation}
where $\Theta_{\mu\nu}$ is given by,
\begin{equation}\label{stressenergy}
\Theta_{\mu\nu}\equiv g^{\alpha\beta}\frac{\delta
T_{\alpha\beta}}{\delta g^{\mu\nu}}
\end{equation}
In the field equations $\nabla_{\mu}$ denotes covariant derivative
associated with the Levi-Civita connection of the metric and
$\Box\equiv \nabla^{\mu}\nabla_{\mu}$ is the D'Alembertian
operator. Moreover we have denoted $f_{R}(R,T)=\partial
f(R,T)/\partial R$ and $f_{T}(R,T)=\partial f(R,T)/\partial T$.
The tensor $\Theta_{\mu\nu}$ can be calculated as,
\begin{equation}\label{theta}
\Theta_{\mu\nu}=-2T_{\mu\nu}+g_{\mu\nu}\mathcal{L}_{m}-2g^{\alpha\beta}\frac{\partial^{2}\mathcal{L}_{m}}{\partial
g^{\mu\nu}\partial g^{\alpha\beta}}
\end{equation}
It is seen that the above tensor depends on the matter lagrangian.
For perfect fluid the above tensor becomes,
\begin{equation}\label{thetaperfect}
\Theta_{\mu\nu}=-2T_{\mu\nu}+pg_{\mu\nu}
\end{equation}

The Vaidya metric in the advanced time coordinate system is given
by,
\begin{equation}\label{vaidya}
ds^{2}=f(t,r)dt^{2}+2dtdr+r^{2}\left(d\theta^{2}+\sin^{2}\theta
d\phi^{2}\right)
\end{equation}
~~~~~where $f(t,r)=-\left(1-\frac{m(t,r)}{r}\right)$ and using the
units $G=c=1$. The total energy momentum tensor of the field
equation (\ref{field}) is given by the following sum,
\begin{equation}\label{energymom}
T_{\mu\nu}=T_{\mu\nu}^{(n)}+T_{\mu\nu}^{(m)}
\end{equation}
where $T_{\mu\nu}^{(n)}$ and $T_{\mu\nu}^{(m)}$ are the
contributions from the Vaidya null radiation and perfect fluid
respectively defined as,
\begin{equation}\label{null}
T_{\mu\nu}^{(n)}=\sigma l_{\mu}l_{\nu}
\end{equation}
and
\begin{equation}\label{fluid}
T_{\mu\nu}^{(m)}=(\rho+p)(l_{\mu}\eta_{\nu}+l_{\nu}\eta_{\mu})+pg_{\mu\nu}
\end{equation}
where $'\rho'$ and $'p'$ are the energy density and pressure for
the perfect fluid and $'\sigma'$ is the energy density
corresponding to Vaidya null radiation. In the co-moving
co-ordinates ($t,r,\theta_{1},\theta_{2},...,\theta_{n}$), the two
eigen vectors of energy-momentum tensor namely $l_{\mu}$ and
$\eta_{\mu}$ are linearly independent future pointing null vectors
having components
\begin{equation}\label{vectors1}
l_{\mu}=(1,0,0,0)~~~~ and~~~~
\eta_{\mu}=\left(\frac{1}{2}\left(1-\frac{m}{r}\right),-1,0,0\right)
\end{equation}
and they satisfy the relations
\begin{equation}\label{vectors2}
l_{\lambda}l^{\lambda}=\eta_{\lambda}\eta^{\lambda}=0,~
l_{\lambda}\eta^{\lambda}=-1
\end{equation}
Therefore, the non-vanishing components of the total
energy-momentum tensor will be as follows
\begin{eqnarray*}
T_{00}=\sigma+\rho\left(1-\frac{m(t,r)}{r}\right),&&
~~T_{01}=-\rho \\
\end{eqnarray*}
\begin{eqnarray}\label{energymomentum}
T_{22}=pr^2, && ~~T_{33}=pr^2 \sin^2\theta
\end{eqnarray}
Here we consider matter in the form of perfect barotropic fluid
given by the equation of state
\begin{equation}\label{pressure}
p=\omega\rho
\end{equation}
where '$\omega$' is the barotropic parameter.

The non-vanishing components of the Ricci tensors are given by,
\begin{eqnarray*}
R_{00}=\frac{\left(m-r\right)m''+2\dot{m}}{2r^{2}},&&
~~~~R_{01}=R_{10}=\frac{m''}{2r} \\
\end{eqnarray*}
\begin{eqnarray}\label{einsteintensors}
R_{22}=m', && ~~R_{33}=m'~ \sin^{2}\theta
\end{eqnarray}
where ~$.$~ and ~$'$~ represents the derivatives with respect to
time coordinate $'t'$ and radial coordinate $'r'$ respectively.
For this system the Ricci scalar becomes,
\begin{equation}\label{ricciscalar}
R=\frac{2m'+rm''}{r^{2}}
\end{equation}
The trace of the energy momentum tensor is calculated as,
\begin{equation}\label{trace}
T=g^{\mu\nu}T_{\mu\nu}=2\left(\omega-1\right)\rho
\end{equation}
The relation between density and mass is considered as
\cite{boer1},
\begin{equation}\label{densitymass}
\rho=n\times m(t,r)
\end{equation}
where $n>0$ is the particle number density.

\vspace{4mm}

\subsection{Field equations}
Now we consider some particular classes of $f(R,T)$ modified
gravity models, which are obtained by some explicit functional
forms of $f(R,T)$. Since the field equations depend on the nature
of matter through the tensor $\Theta_{\mu\nu}$, here we will
consider the field equations for a perfect fluid source, which
will be our field of interest in this study, as discussed in the
previous section. On a broad sense we are going to discuss two
types of models.

\subsubsection{Model-1:  $f(R,T)=f_{1}(R)+f_{2}(T)$}
Here we consider models of the form $f(R,T)=f_{1}(R)+f_{2}(T)$,
where $f_{1}(R)$ and $f_{2}(T)$ are arbitrary functions of $R$ and
$T$ respectively. It is straightforward to see that for
$f_{1}(R)=R$ and $f_{2}(T)=0$, we can retrieve GR from this model.
Using Eq.(\ref{field}), the gravitational field equations for this
model is given by,
\begin{equation}\label{fieldmodel1}
f_{1}'(R)R_{\mu\nu}-\frac{1}{2}f_{1}(R)g_{\mu\nu}+\left(g_{\mu\nu}\Box-\nabla_{\mu}\nabla_{\nu}\right)f_{1}'(R)=\kappa
T_{\mu\nu}+f_{2}'(T)T_{\mu\nu}+\left[f_{2}'(T)p+\frac{1}{2}f_{2}(T)\right]g_{\mu\nu}
\end{equation}
where $'$ represents derivative with respect to the argument. Now
using the Vaidya metric given in Eq.(\ref{vaidya}) and using
Eqs.(\ref{energymomentum}), (\ref{pressure}),
(\ref{einsteintensors}) and (\ref{ricciscalar}) in
Eq.(\ref{fieldmodel1}), we compute all the components of the
Einstein's field equations for this model (taking $\kappa=1$).
Here we report the $(01)$, $(22)$ and $(33)$ components of the
field equations which will be used in our analysis. The rest of
the components are reported in the appendix
section of the paper.\\

\textbf{\textit{1. The $(01)$-component of field equations is
given by,}}
\begin{equation}\label{fieldeq01}
-r\left\{f_{1}(R)+f_{2}(T)-2\rho\left(1+f_{2}'(T)-\omega
f_{2}'(T)\right)\right\}+f_{1}'(R)m''=0
\end{equation}

\textbf{\textit{2. The $(22)$ and $(33)$ components of field
equations are given by,}}
\begin{equation}\label{fieldeq2233}
r^{2}\left[f_{1}(R)+f_{2}(T)+2\omega \rho+4f_{2}'(T)\omega
\rho\right]-2f_{1}'(R)m'=0
\end{equation}
The above equations along with the ones reported in the appendix
are the Einstein's field equations for $f(R,T)$ gravity in the
time dependent Vaidya spacetime for the first model.

\subsubsection{Model-2:  $f(R,T)=f_{1}(R)+f_{2}(R)f_{3}(T)$}
Now we consider a second model given by
$f(R,T)=f_{1}(R)+f_{2}(R)f_{3}(T)$, where $f_{i}(R)$, $i=1, 2$ are
arbitrary functions of $R$ and $f_{3}(T)$ is an arbitrary function
of $T$. Here the scalar invariants $R$ and $T$ are non-minimally
coupled to each other via the second term. In order to realize GR
from this model, we should have $f_{1}(R)=R$ and either or both of
$f_{2}(R)$ and $f_{3}(T)$ equal to zero. We may also take
$f_{1}(R)=0, f_{2}(R)=R$ and $f_{3}(T)=1$ to get GR from this
model. Using Eq.(\ref{field}), the gravitational field equations
for this model is given by,

$$R_{\mu\nu}\left[f_{1}'(R)+f_{2}'(R)f_{3}(T)\right]-\frac{1}{2}f_{1}(R)g_{\mu\nu}+\left(g_{\mu\nu}\Box-\nabla_{\mu}\nabla_{\nu}\right)\left[f_{1}'(R)+f_{2}'(R)f_{3}(T)\right]$$
\begin{equation}\label{fieldmodel1}
=\kappa
T_{\mu\nu}+f_{2}(R)f_{3}'(T)T_{\mu\nu}+f_{2}(R)\left[f_{3}'(T)p+\frac{1}{2}f_{3}(T)\right]g_{\mu\nu}
\end{equation}

Like the previous model here also we report the necessary
components of the field equations for the second model (taking $\kappa=1$):\\

\textbf{\textit{1. The $(01)$-component of field equations is
given by,}}
\begin{equation}\label{fieldeq01m2}
-r\left[f_{1}(R)-2\rho+f_{2}(R)\left\{f_{3}(T)+2f_{3}'(T)\rho\left(\omega-1\right)\right\}\right]+
\left(f_{1}'(R)+f_{2}'(R)f_{3}(T)\right)m''=0
\end{equation}

\textbf{\textit{2. The $(22)$ and $(33)$ components of field
equations are given by,}}
\begin{equation}\label{fieldeq22m2}
r^{2}\left(f_{1}(R)+f_{2}(R)f_{3}(T)+2\omega\rho+4f_{2}(R)f_{3}'(T)\omega\rho\right)
-2\left(f_{1}'(R)+f_{2}'(R)f_{3}(T)\right)m'=0
\end{equation}

The rest of the components are reported in the appendix section.
The above equations along with the ones reported in the appendix
are the Einstein's field equations for $f(R,T)$ gravity in the
time dependent Vaidya spacetime for the second model.

\subsection{Solution of the system}
In this section we will proceed to find solutions of the above
systems. In order to do that, we will have to consider special
forms for the arbitrary functions of $R$ and $T$ as examples. We
will do this for both the models studied in the previous section.

\subsubsection{Model-1}
Here we have consider four different toy models as examples in
order to solve the field equations. The model involves arbitrary
functions of $R$ and $T$ coupled in a minimal way. The functional
forms used in the toy models are basically power law and
exponential forms, which are observationally the most favoured
models with constraints imposed on their free parameters.\\

\textbf{Case-1: $f_{1}(R)=g_{1}R^{\beta_{1}},~~
f_{2}(T)=g_{2}T^{\beta_{2}}$}, where $g_{1}, \beta_{1}, g_{2}, \beta_{2}$ are constants.\\\\
Here we have chosen power law forms for both $f_{1}(R)$ and
$f_{2}(T)$. For our convenience we call this the double-power (DP)
model. For $g_{2}=0$ and $g_{1}=\beta_{1}=1$, we get back GR from
this model. For this case, the $22$ or $33$ component of the field
equations give the following differential equation,

$$\frac{1}{r^{2}}\left[g_{1}\left\{2\left(\beta_{1}-1\right)m'-rm''\right\}
\left\{\frac{2m'+rm''}{r^{2}}\right\}^{\beta_{1}-1}\right]
+\frac{1}{1-\omega}\left[2^{\beta^{2}}g_{2}\left(2\beta_{2}\omega+\omega-1\right)
\left\{n\left(\omega-1\right)m\right\}^{\beta_{2}}\right]$$
\begin{equation}\label{diffeq1}
-2n\omega m=0
\end{equation}

Solving the above differential equation for
$\beta_{1}=\beta_{2}=1$ we get,
\begin{equation}\label{solution1}
m(t,r)=h_{1}(t)~AiryAi\left[2^{1/3}r\left\{\frac{n\left(g_{2}-\omega-3g_{2}\omega\right)}
{g_{1}}\right\}^{1/3}\right]+h_{2}(t)~AiryBi\left[2^{1/3}r\left\{\frac{n\left(g_{2}-\omega-3g_{2}\omega\right)}
{g_{1}}\right\}^{1/3}\right]
\end{equation}
where $AiryAi$ and $AiryBi$ are the two Airy functions (see
appendix) and $h_{1}(t)$, $h_{2}(t)$ are arbitrary functions of
time which arises from integration. We would like to mention here
that the imposed conditions $\beta_{1}=\beta_{2}=1$ are necessary
to get a solution of this system by the known mathematical
methods.

Now the $01$ component of the field equations gives the
differential equation,
\begin{equation}\label{diffeq2}
2nr^{2}m-2^{\beta_{2}}g_{2}r^{2}\left(1+\beta_{2}\right)\left\{n\left(\omega-1\right)m\right\}
^{\beta_{2}}+g_{1}\left(\frac{2m'+rm''}{r^2}\right)^{\beta_{1}-1}\left\{r\left(\beta_{1}-1\right)
m''-2m'\right\}=0
\end{equation}
Solving this equation for $\beta_{1}=\beta_{2}=1$ we get,
\begin{equation}\label{solution2}
m(t,r)=h_{3}(t)e^{\frac{nr^{3}\left(1-2g_{2}\left(\omega-1\right)\right)}{3g_{1}}}
\end{equation}
where $h_{3}(t)$ is an arbitrary function of time. For some values
of the arbitrary functions it is expected that the solutions given
by Eqs.(\ref{solution1}) and (\ref{solution2}) will match. In fact
it has been checked that both these solutions give rise to similar
scenarios in the gravitational collapse scheme which we are going
to introduce in the next section. So we are going adopt one of
these solutions depending upon the nature of genericity of the
solution. Since the solution given by Eq.(\ref{solution1}) has two
arbitrary functions we will use this for our collapse study simply
because it is more general in nature and can easily generate the
other solution for some well chosen initial conditions. From here
on, we will only concentrate on the differential equation and the
solution obtained from the $22$ or $33$ component of the field
equations for the reason discussed above. Now that we have
obtained the mass parameter, using it in the Vaidya metric given
in Eq.(\ref{vaidya}), we can easily get the Vaidya spacetime for
the corresponding model in $f(R,T)$ gravity.
\\

\textbf{Case-2: $f_{1}(R)=g_{1}e^{\beta_{1}R},~~
f_{2}(T)=g_{2}e^{\beta_{2}T}$}, where $g_{1}, \beta_{1}, g_{2}, \beta_{2}$ are
constants.\\\\
Here we have chosen exponential forms for both $f_{1}(R)$ and
$f_{2}(T)$. For our convenience we call this the
double-exponential (DE) model. Realizing GR from this model is
difficult. Nevertheless an approximation will help us realize the
scenario. Expanding $e^{\beta_{1}R}$ in Taylor's series and
keeping the linear terms in $R$ only, will help us realize GR for
$g_{2}=0, g_{1}=1$ and $\beta_{1}=\frac{R-1}{R}$. For this case,
the $22$ or $33$ component of the field equations gives us the
differential equation,
\begin{equation}\label{diffeq3}
2nr^{2}\omega\left\{1+2g_{2}\beta_{2}e^{2n\beta_{2}\left(\omega-1\right)m}\right\}m
+e^{\frac{\beta_{1}\left(2m'+rm''\right)}{r^{2}}}g_{1}\left(r^{2}
-2\beta_{1}m'\right)+g_{2}r^{2}e^{2n\beta_{2}\left(\omega-1\right)m}=0
\end{equation}
This equation has got the unknown function $m$ and its derivatives
in exponential form. It is not possible to find a general solution
of this equation. So we search for approximate solutions. We
expand the exponentials in the first and the third term in Taylor
series and take the linear terms only to get the following
solution for $\beta_{1}=0$,
\begin{equation}\label{solution3}
m(t,r)=\frac{g_{2}n\beta_{2}\left(1-3\omega\right)-n\omega+\sqrt{n^{2}\left[\omega^{2}
+g_{2}^{2}\beta_{2}^{2}\left(1+\omega\right)^{2}+2g_{2}\beta_{2}\omega
\left\{3\omega-4g_{1}\beta_{2}\left(\omega-1\right)-1\right\}\right]}}
{8g_{2}n^{2}\beta_{2}^{2}\omega\left(\omega-1\right)}
\end{equation}
We see that this a constant solution for the mass parameter. Using
this in Eq.(\ref{vaidya})
we will get the Vaidya spacetime in $f(R,T)$ gravity for this case.\\

\textbf{Case-3: $f_{1}(R)=g_{1}R^{\beta_{1}},~~
f_{2}(T)=g_{2}e^{\beta_{2}T}$}, where $g_{1}, \beta_{1}, g_{2},
\beta_{2}$ are
constants.\\\\
Here we have chosen power law for $f_{1}(R)$ and exponential form
for $f_{2}(T)$. For our convenience we call this the
power-exponential (PE) model. For $g_{2}=0$ and
$g_{1}=\beta_{1}=1$, we get back GR from this model. The $22$ or
$33$ component of the field equations gives us the differential
equation,
\begin{equation}\label{diffeq4}
2nr^{2}\left(1+2g_{2}\beta_{2}e^{2n\beta_{2}\left(\omega-1\right)m}\right)\omega
m+e^{2n\beta_{2}\left(\omega-1\right)m}g_{2}r^{2}-g_{1}\left\{2\left(\beta_{1}-1\right)m'
-rm''\right\}\left(\frac{2m'+rm''}{r^{2}}\right)^{\beta_{1}-1}=0
\end{equation}
For $\beta_{1}=1$ and $\beta_{2}=0$ we get the following solution
for the above differential equation,

$$m(t,r)=\frac{1}{2n\omega}\left[g_{2}\pi
AiryAi'\left[2^{1/3}r\left(-\frac{n\omega}{g_{1}}\right)^{1/3}\right]AiryBi\left[-\frac{2^{1/3}nr\omega}
{g_{1}\left(-\frac{n\omega}{g_{1}}\right)^{2/3}}\right]-g_{2}\pi
AiryAi\left[-\frac{2^{1/3}nr\omega}
{g_{1}\left(-\frac{n\omega}{g_{1}}\right)^{2/3}}\right]\times\right.$$

\begin{equation}\label{solution4}
\left.AiryBi'\left[2^{1/3}r\left(-\frac{n\omega}{g_{1}}\right)
^{1/3}\right]\right]+h_{4}(t)AiryAi\left[-\frac{2^{1/3}nr\omega}
{g_{1}\left(-\frac{n\omega}{g_{1}}\right)^{2/3}}\right]+h_{5}(t)AiryBi\left[-\frac{2^{1/3}nr\omega}
{g_{1}\left(-\frac{n\omega}{g_{1}}\right)^{2/3}}\right]
\end{equation}
where $AiryAi'$, $AiryBi'$ are derivatives of the Airy functions
with respect to the argument and $h_{4}(t)$, $h_{5}(t)$ are
arbitrary functions of time.
\\\\

\textbf{Case-4: $f_{1}(R)=g_{1}e^{\beta_{1}R},~~
f_{2}(T)=g_{2}T^{\beta_{2}}$}, where $g_{1}, \beta_{1}, g_{2}, \beta_{2}$ are
constants.\\\\
Here we have chosen exponential form for both $f_{1}(R)$ and power
law for $f_{2}(T)$. For our convenience we call this the
exponential-power (EP) model. A similar scenario as discussed in
case-2, will help us realize GR from this model. The $22$ or $33$
component of the field equations gives us the differential
equation,
\begin{equation}\label{diffeq5}
r^{2}\left[-2n\omega
m-\frac{2^{\beta_{2}}g_{2}\left(\omega+2\beta_{2}\omega-1\right)\left\{n\left(\omega-1\right)m
\right\}^{\beta_{2}}}{\omega-1}\right]-g_{1}\left(r^{2}-2\beta_{1}m'\right)
e^{\frac{\beta_{1}\left(2m'+rm''\right)}{r^{2}}}=0
\end{equation}
For $\beta_{1}=0$ and $\beta_{2}=1$ we get the following solution
of the above equation,
\begin{equation}\label{solution5}
m(t,r)=-\frac{g_{1}}{2n\left(\omega+3g_{2}\omega-g_{2}\right)}
\end{equation}
\\

\subsubsection{Model-2}

Now again we consider some special models as sub-cases in order to
solve the field equations. The basic difference between this model
with the previous one is that here the functions of $R$ and $T$
will be minimally coupled to each other which is observationally
the more favoured model.\\

\textbf{Case-1: $f_{1}(R)=g_{1}R^{\beta_{1}},~
f_{2}(R)=g_{2}R^{\beta_{2}},~ f_{3}(T)=g_{3}T^{\beta_{3}}$}~~~~~~($g_{1}, \beta_{1}, g_{2}, \beta_{2}, g_{3}, \beta_{3}$ are constants)\\

Here we have considered power law forms for all the three
functions. We call this the triple-power (TP) model. For
$g_{1}=\beta_{1}=1$ and $g_{2}=0$ or $g_{3}=0$, we can realize GR
from this model. We may also realize GR for $g_{1}=\beta_{3}=0$
and $g_{2}=g_{3}=\beta_{2}=1$. The $22$ or $33$ component of the
field equations gives us the differential equation,

$$2n\omega
m\left(2m'+rm''\right)+g_{1}\left(\frac{2m'+rm''}{r^{2}}\right)^{\beta_{1}}
\left\{2\left(1-\beta_{1}\right)m'+rm''\right\}$$
\begin{equation}\label{diffeq6}
+\frac{2^{\beta_{3}}g_{2}g_{3}
\left\{n\left(\omega-1\right)m\right\}^{\beta_{3}}\left(\frac{2m'+rm''}{r^{2}}\right)
^{\beta_{2}}\left\{2\left(\beta_{2}+\omega-\beta_{2}\omega+2\beta_{3}\omega-1\right)m'
+r\left(\omega+2\beta_{3}\omega-1\right)m''\right\}}{\omega-1}=0
\end{equation}
A solution for the above differential equation can be obtained for
$\beta_{1}=1, \beta_{3}=2$ and $\omega=1$ which is given below,
\begin{equation}\label{solution6}
m(t,r)=h_{6}(t)AiryAi\left[-\frac{2^{1/3}nr}{g_{1}\left(-n/g_{1}\right)^{2/3}}\right]
+h_{7}(t)AiryBi\left[-\frac{2^{1/3}nr}{g_{1}\left(-n/g_{1}\right)^{2/3}}\right]
\end{equation}
where $h_{6}(t)$ and $h_{7}(t)$ are arbitrary functions of time.
This solution corresponds to early universe ($\omega=1$)
representing stiff perfect fluid.
\\\\

\textbf{Case-2: $f_{1}(R)=g_{1}e^{\beta_{1}R},~~
f_{2}(R)=g_{2}e^{\beta_{2}R},~~ f_{3}(T)=g_{3}T^{\beta_{3}}$}~~($g_{1}, \beta_{1}, g_{2}, \beta_{2}, g_{3}, \beta_{3}$ are constants)\\

This is the double-exponential-power (DEP) model. For this model
the $22$ or $33$ component of the field equations yields the
following differential equation,

$$2nr^{2}\left(\omega-1\right)\omega
m+g_{1}\left(\omega-1\right)\left(r^{2}e^{\frac{\beta_{1}\left(2m'+rm''\right)}{r^{2}}}
-2\beta_{1}e^{\beta_{1}R}m'\right)+2^{\beta_{3}}g_{2}g_{3}\left\{n\left(\omega-1\right)m\right\}
^{\beta_{3}}\times$$
\begin{equation}\label{diffeq7}
\left\{r^{2}\left(\omega+3\beta_{3}\omega-1\right)e^{\frac{\beta_{2}
\left(2m'+rm''\right)}{r^{2}}}-2\beta_{2}\left(\omega-1\right)e^{\beta_{2}R}m'\right\}=0
\end{equation}
For $\beta_{1}=0, \beta_{3}=2$ and $\omega=1$ we get the following
constant solution of the above equation,
\begin{equation}\label{solution7}
m(t,r)=-\frac{g_{1}}{2n}
\end{equation}
We see that this is a constant solution. Moreover this solution is
valid in the early universe for a stiff perfect fluid
($\omega=1$).
\\\\

\textbf{Case-3: $f_{1}(R)=g_{1}R^{\beta_{1}},~~
f_{2}(R)=g_{2}R^{\beta_{2}},~~ f_{3}(T)=g_{3}e^{\beta_{3}T}$}~~($g_{1}, \beta_{1}, g_{2}, \beta_{2}, g_{3}, \beta_{3}$ are constants)\\

From the choice of the functions we can see that this is a
double-power-exponential (DPE) model. For $g_{2}=0$ or $g_{3}=0$
and $g_{1}=\beta_{1}=1$ we get back GR from this model. In this
case the $(22)$ or $(33)$ components of the field equations gives
the differential equation,

$$r^{2}\left[2n\omega
m+g_{1}\left(\frac{2m'+rm''}{r^{2}}\right)^{\beta_{1}}+g_{2}g_{3}\left(1+4n\beta_{3}\omega
m\right)e^{2n\beta_{3}\left(\omega-1\right)m}\left(\frac{2m'+rm''}{r^{2}}\right)^{\beta_{2}}\right]$$
\begin{equation}\label{diffeq8}
-2m'\left[g_{1}\beta_{1}\left(\frac{2m'+rm''}{r^{2}}\right)^{\beta_{1}-1}
+g_{2}g_{3}\beta_{2}e^{2n\beta_{3}\left(\omega-1\right)m}\left(\frac{2m'+rm''}{r^{2}}\right)
^{\beta_{2}-1}\right]=0
\end{equation}
The following solution for the above equation is obtained for
$\beta_{1}=\beta_{2}=1$ and $\omega=0$,
\begin{equation}\label{solution8}
m(t,r)=h_{8}(t)+h_{9}(t)r ~~~~~~~~ OR
~~~~~~m(t,r)=\frac{\log\left(-\frac{g_{2}g_{3}}{g_{1}}\right)}{2n\beta_{3}}
\end{equation}
where $h_{8}(t)$ and $h_{9}(t)$ are arbitrary functions of time.
We will use the first expression for the mass parameter for
further study because it is evolving with $r$ and $t$ and hence is
more informative for our study. Moreover the second expression
being a constant can always be realized from the first expression
using suitable initial conditions. In this sense the first
expression is more generalized and so we intend to use it in our
analysis.\\\\

\textbf{Case-4: $f_{1}(R)=g_{1}e^{\beta_{1}R},~~
f_{2}(R)=g_{2}e^{\beta_{2}R},~~ f_{3}(T)=g_{3}e^{\beta_{3}T}$}~~($g_{1}, \beta_{1}, g_{2}, \beta_{2}, g_{3}, \beta_{3}$ are constants)\\

This is the triple-exponential (TE) model formed by three
exponential functions. For this model the $(22)$ or $(33)$
components of the field equations give the differential equation,
\begin{equation}\label{diffeq9}
g_{1}r^{2}e^{\frac{\beta_{1}\left(2m'+rm''\right)}{r^{2}}}+2nr^{2}\omega
m-2g_{1}\beta_{1}e^{\frac{\beta_{1}\left(2m'+rm''\right)}{r^{2}}}m'+g_{2}g_{3}
\left(r^{2}+4nr^{2}\beta_{3}\omega
m-2\beta_{2}m'\right)e^{2n\beta_{3}\left(\omega-1\right)m
+\frac{\beta_{2}\left(2m'+rm''\right)}{r^{2}}}=0
\end{equation}
A solution to the above equation is obtained for
$\beta_{1}=\beta_{2}=1$ and $\omega=0$ which is given below,
\begin{equation}\label{solution9}
m(t,r)=\frac{r^{3}}{6}+h_{10}(t)  ~~~~~~~OR
~~~~~~~m(t,r)=\frac{\log\left(-\frac{g_{2}g_{3}}{g_{1}}\right)}{2n\beta_{3}}
\end{equation}
where $h_{10}(t)$ is an arbitrary function of time. Just like the
previous model, here also we will use the first expression for the
mass parameter, for reasons similar to the ones discussed in the
previous model. It should be noted that this solution corresponds
to dust ($\omega=0$) as far as the matter content of the universe
is concerned and cosmologically this corresponds to early
universe.

\section{Gravitational Collapse}
In this section, we will devise a mechanism in order to study the
gravitational collapse of a massive star in this system. As
mentioned before, we will consider that the parent star is a
massive one so that collapse smoothly continues until a
singularity (BH or NS) is formed. Our idea is to develop a set-up,
via which the nature of the singularity (BH or NS) can be
comprehensively identified. At least our aim is to derive a
condition that will govern the nature of singularity (BH or NS)
resulting out of the gravitational collapse.

Let us consider a spherical collapsing system, where the physical
radius of the $r$-th shell of the star at time $t$ is $R(t,r)$. A
suitable initial condition would be that in the epoch $t=0$, we
have $R(0,r)=r$. It is obvious that if the collapse is
inhomogeneous, then different collapsing shells may become
singular at different times. We are concerned with the light
photons emerging from the singularity and travelling along the
geodesics and reaching an external observer. An event horizon will
be an obstruction for these photons and will resist them from
reaching the observer. So here we will probe the existence of such
outgoing non-spacelike geodesics. Theoretically if such geodesics
possess well defined tangent at the singularity, the quantity
$dR/dr$ will definitely tend towards a finite limit with the
geodesics approaching the singularity in the past following the
trajectories. When these trajectories reach the points $(t_{0},
r)=(t_{0}, 0)$, there is a complete breakdown of mathematical and
physical concepts and a singularity occurs at $R(t_{0}, 0)=0$. At
these points ideally the collapsing matter shells are crushed to
zero radius, which results in the formation of the central
singularity. This is a highly compact object, since a huge amount
of mass is packed inside an almost negligible volume. Now if we
follow back the path of the outgoing non-spacelike geodesics that
are emerging from the central singularity, it is highly probable
that they will terminate in the past at the singularity $(r=0,
t=t_0)$ where $R(t_{0},0)=0$. Therefore from our set-up,
mathematically we should have $R\rightarrow 0$ as $r\rightarrow 0$
\cite{sing}.

We obtain the equation for outgoing radial null geodesics from the
Vaidya metric (\ref{vaidya}) by putting $ds^{2}=0$ and
$d\Omega_{2}^{2}=d\theta^{2}+\sin^{2}\theta d\phi^{2}=0$ as
furnished below
\begin{equation}
\frac{dt}{dr}=\frac{2}{\left(1-\frac{m(t,r)}{r}\right)}.
\end{equation}
The above differential equation has a singularity at $r=0,~t=0$.
Mathematically this means that any solution to the above equation
is not analytic at the point $r=0,~t=0$. Since there is a
mathematical breakdown at the singularity we are forced to study
the limiting behaviour as one approaches the singularity. To
facilitate this, we consider a parameter $X=t/r$. The idea is to
study the limiting behaviour of the function $X$ as we approach
the singularity at $r=0,~t=0$ following the radial null geodesic.
If we denote the limiting value of $X$ by $X_{0}$ then using
L'Hospital's rule we have
\begin{eqnarray}\label{X0}
\begin{array}{c}
X_{0}\\\\
{}
\end{array}
\begin{array}{c}
=\lim X \\
\begin{tiny}t\rightarrow 0\end{tiny}\\
\begin{tiny}r\rightarrow 0\end{tiny}
\end{array}
\begin{array}{c}
=\lim \frac{t}{r} \\
\begin{tiny}t\rightarrow 0\end{tiny}\\
\begin{tiny}r\rightarrow 0\end{tiny}
\end{array}
\begin{array}{c}
=\lim \frac{dt}{dr} \\
\begin{tiny}t\rightarrow 0\end{tiny}\\
\begin{tiny}r\rightarrow 0\end{tiny}
\end{array}
\begin{array}{c}
=\lim \frac{2}{\left(1-\frac{m(t,r)}{r}\right)} \\
\begin{tiny}t\rightarrow 0\end{tiny}~~~~~~~~~~~~\\
\begin{tiny}r\rightarrow 0\end{tiny}~~~~~~~~~~~~
 {}
\end{array}
\end{eqnarray}
This will actually generate an algebraic equation in terms of
$X_{0}$. The roots of this equation will be our prime concern
because they actually represent the slopes (direction) of the
tangents to the geodesics. Here we are only interested in the real
roots because we are dealing with a realistic collapsing scenario
with no connection to the complex domain. For our set-up, any
positive real root of this algebraic equation will give the
direction of the tangent to an outgoing null geodesic at the
singularity. Therefore the existence of positive real roots of
this equation corresponds to a necessary and sufficient condition
for the singularity to be naked in nature. Now as we have
discussed earlier, if a single null geodesic in the $(t,r)$ plane
escapes the singularity, it would mean that a single wavefront
emitted from the singularity reaches the external observer. In
such a scenario the singularity would be visible only
instantaneously to a distant observer and become a \textit{locally
naked singularity}. Physically this will correspond to a situation
where the event horizon was eliminated from the picture, but only
temporarily. But this might not be enough for a complete exchange
of information between the singularity and the observer. So for a
formidable exchange of information, the singularity is to be seen
for a finite period of time. This requires a family of null
geodesics escaping from the singularity thus making it
\textit{globally naked}. In our mathematical set-up this can be
investigated very easily from the number of real positive roots
obtained from the above algebraic equation. The above explained
comprehensive mathematical set-up for identifying the nature of
singularity formed as the end state of a gravitational collapse
was first used by Joshi, Singh and Dwivedi in several of their
papers \cite{cch2, cch3, sing, ns5}. With the mathematical tools
ready, we proceed to study the models one by one.

\subsection{Model-1}

\subsubsection{Case-1}
Using Eq.(\ref{solution1}) in Eq.(\ref{X0}) we get,

$$\frac{2}{X_{0}}=
\begin{array}llim\\
\begin{tiny}t\rightarrow 0\end{tiny}\\
\begin{tiny}r\rightarrow 0\end{tiny}
\end{array}
\left[1-\frac{h_{1}(t)}{r}~AiryAi\left[2^{1/3}r\left\{\frac{n\left(g_{2}-\omega-3g_{2}\omega\right)}
{g_{1}}\right\}^{1/3}\right]\right.$$
\begin{eqnarray}\label{X01}
\left.-\frac{h_{2}(t)}{r}~AiryBi\left[2^{1/3}r\left\{\frac{n\left(g_{2}-\omega-3g_{2}\omega\right)}
{g_{1}}\right\}^{1/3}\right]\right]
\end{eqnarray}
Here we will consider self-similar collapsing scenario. So we
consider the following self-similar expressions for the arbitrary
functions $h_{i}(t)$, $i=1,2$\\\\
$h_{1}(t)=\xi_{1}t, ~~~~~~~~   h_{2}(t)=\xi_{2}t$ ~~~~~~~ where
$\xi_{1}$ and $\xi_{2}$ are arbitrary constants.\\\\ Using the
above chosen functions in Eq.(\ref{X01}) we get the following
algebraic equation in $X_{0}$,
\begin{equation}\label{algebraic1}
\frac{1}{\Gamma(2/3)}\left(\frac{\xi_{1}}{3^{2/3}}+\frac{\xi_{2}}{3^{1/6}}\right)X_{0}^{2}-X_{0}+2=0
\end{equation}
To evaluate the above limit, we have used the values of the Airy
functions given by Eq.(\ref{airy0}) in the appendix. Solving the
above equation we get two values of $X_{0}$ which are,
\begin{equation}\label{X0sol1}
X_{0_{1,2}}^{case1}=\frac{\sqrt{\Gamma(2/3)}}{2\sqrt{3}\left(\frac{\xi_{1}}{3^{2/3}}+\frac{\xi_{2}}{3^{1/6}}
\right)}\left[\sqrt{3\Gamma(2/3)}\pm\sqrt{3\Gamma(2/3)-8\times
3^{1/3}\xi_{1}-8\times 3^{5/6}\xi_{2}}\right]
\end{equation}
Here we have considered and henceforth we will consider positive
sign for root1 and negative sign for root2. Since we are dealing
with a realistic situation we should have
$3^{1/3}\xi_{1}+3^{5/6}\xi_{2}\leq \frac{3\Gamma(2/3)}{8}$. Now in
order to get a NS we should have
$X_{0}>0$. We list below the respective conditions in detail.\\

Conditions for a local NS:~~~~~  $X_{0_{1}}^{case1}>0$~~ $\&$~~  $X_{0_{2}}^{case1}<0$ ~~~~~OR~~~~~$X_{0_{1}}^{case1}<0$~~ $\&$~~  $X_{0_{2}}^{case1}>0$\\

Conditions for global NS:~~~~~~   $X_{0_{1}}^{case1}>0$ ~~ $\&$~~ $X_{0_{2}}^{case1}>0$\\

Condition for BH:~~~~~~~~~~~~~~~~   $X_{0_{1}}^{case1}<0$ ~~ $\&$~~
$X_{0_{2}}^{case1}<0$\\\\
We see that the above conditions put constraints of $\xi_{1}$ and
$\xi_2$. So by clubbing this theory with observations of
collapsing massive stars, we can get bounds on the model
parameters. The roots $X_{0_{1,2}}^{case1}$ have been plotted against the parameters $\xi_{1}$ and $\xi_{2}$ in Figs.(1) and (2).\\\\

\begin{figure}\label{f1}
~~~~~~~~~\includegraphics[height=1.7in,width=2.5in]{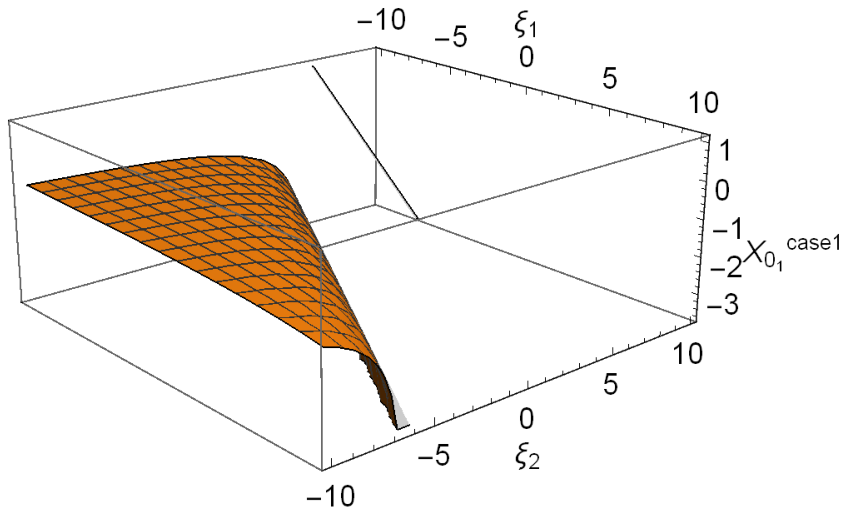}~~~~~~~~~~~~~~~~~~\includegraphics[height=1.7in,width=2.5in]{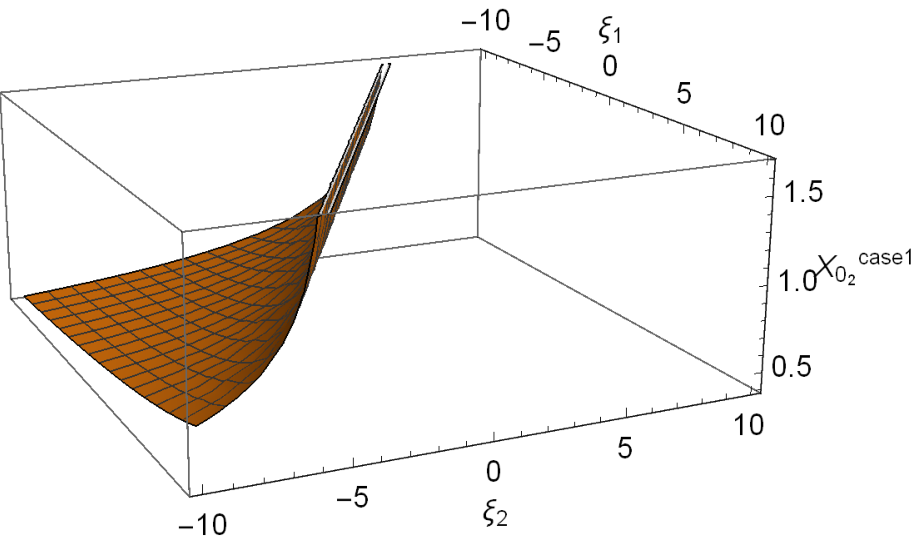}~~~~~~~\\

~~~~~~~~~~~~~~~~~~~~~~~~~~~~Fig.1~~~~~~~~~~~~~~~~~~~~~~~~~~~~~~~~~~~~~~~~~~~~~~~~~~~~~~~~~~~~~Fig.2~~~~~~~~~\\

\vspace{1mm} \textit{\textbf{Figs.1 and 2} show the variation of
the collapse parameter $X_{0}$ for different values of $\xi_1$ and
$\xi_2$ for Case-1 of Model-1. Fig.1 shows the variation for the
first root $X_{0_1}^{case1}$, whereas Fig.2 shows the variation
for the second root $X_{0_2}^{case1}$.}
\end{figure}

\subsubsection{Case-2}
Using Eq.(\ref{solution3}) in Eq.(\ref{X0}) we get,

$$\frac{2}{X_{0}}=
\begin{array}llim\\
\begin{tiny}t\rightarrow 0\end{tiny}\\
\begin{tiny}r\rightarrow 0\end{tiny}
\end{array}
\left[1-\frac{g_{2}n\beta_{2}\left(1-3\omega\right)-n\omega+\sqrt{n^{2}\left[\omega^{2}
+g_{2}^{2}\beta_{2}^{2}\left(1+\omega\right)^{2}+2g_{2}\beta_{2}\omega
\left\{3\omega-4g_{1}\beta_{2}\left(\omega-1\right)-1\right\}\right]}}
{8g_{2}n^{2}\beta_{2}^{2}\omega\left(\omega-1\right)r}\right]$$
\begin{eqnarray}\label{X02}
.
\end{eqnarray}
Evaluating the above limit we get $\frac{2}{X_{0}}\rightarrow
\infty$, which implies $X_{0}\rightarrow 0$. Since the mass
function in this case is not a function of $t$ and $r$, we do not
get a realistic collapsing scenario for this particular model
according to our scheme of study.

\subsubsection{Case-3}
Using Eq.(\ref{solution4}) in Eq.(\ref{X0}) we get,
$$\frac{2}{X_{0}}=
\begin{array}llim\\
\begin{tiny}t\rightarrow 0\end{tiny}\\
\begin{tiny}r\rightarrow 0\end{tiny}
\end{array}
\left[1-\frac{1}{2n\omega r}\left\{g_{2}\pi
AiryAi'\left[2^{1/3}r\left(-\frac{n\omega}{g_{1}}\right)^{1/3}\right]\times
AiryBi\left[-\frac{2^{1/3}nr\omega}
{g_{1}\left(-\frac{n\omega}{g_{1}}\right)^{2/3}}\right]\right.\right.$$

$$\left.\left.-g_{2}\pi AiryAi\left[-\frac{2^{1/3}nr\omega}
{g_{1}\left(-\frac{n\omega}{g_{1}}\right)^{2/3}}\right]\times
AiryBi'\left[2^{1/3}r\left(-\frac{n\omega}{g_{1}}\right)
^{1/3}\right]\right\}-\frac{h_{4}(t)}{r}AiryAi\left[-\frac{2^{1/3}nr\omega}
{g_{1}\left(-\frac{n\omega}{g_{1}}\right)^{2/3}}\right]\right.$$

\begin{eqnarray}\label{X03}
\left.-\frac{h_{5}(t)}{r}AiryBi\left[-\frac{2^{1/3}nr\omega}
{g_{1}\left(-\frac{n\omega}{g_{1}}\right)^{2/3}}\right]\right]
\end{eqnarray}
We consider the following functions: $h_{4}(t)=\xi_{4}t, ~~
h_{5}(t)=\xi_{5}t$, ~ where $\xi_{4}$ and $\xi_{5}$ are arbitrary
constants. Using the above chosen functions in Eq.(\ref{X03}) we
get the following algebraic equation in $X_{0}$,
\begin{equation}\label{algebraic2}
\frac{1}{\Gamma(2/3)}\left(\frac{\xi_{4}}{3^{2/3}}+\frac{\xi_{5}}{3^{1/6}}\right)X_{0}^{2}+\left(\xi_{3}-1\right)X_{0}+2=0
\end{equation}
where $\xi_{3}$ is a constant that arises as a limiting value of
the second term of the expression in Eq.(\ref{X02}). Solving the
above quadratic we get,
\begin{equation}\label{X0sol2}
X_{0_{1,2}}^{case3}=\frac{\Gamma(2/3)\left[1-\xi_{3}\pm
\sqrt{\left(\xi_{3}-1\right)^{2}-\frac{8}{\Gamma(2/3)}\left(\frac{\xi_{4}}{3^{2/3}}+\frac{\xi_{5}}{3^{1/6}}\right)}\right]}{2\left(\frac{\xi_{4}}{3^{2/3}}+\frac{\xi_{5}}{3^{1/6}}\right)}
\end{equation}
where we should have
$\left(\xi_{3}-1\right)^{2}\geq\frac{8}{\Gamma(2/3)}\left(\frac{\xi_{4}}{3^{2/3}}+\frac{\xi_{5}}{3^{1/6}}\right)$.
The conditions for NS or BH will be similar as discussed in Case-1.\\

Conditions for a local NS:~~~~~  $X_{0_{1}}^{case3}>0$~~ $\&$~~  $X_{0_{2}}^{case3}<0$~~~~~OR~~~~~$X_{0_{1}}^{case3}<0$~~ $\&$~~  $X_{0_{2}}^{case3}>0$\\

Conditions for global NS:~~~~~~   $X_{0_{1}}^{case3}>0$ ~~ $\&$~~ $X_{0_{2}}^{case3}>0$\\

Condition for BH:~~~~~~~~~~~~~~~~   $X_{0_{1}}^{case3}<0$ ~~$\&$~~$X_{0_{2}}^{case3}<0$\\\\
We see that the above conditions put numerical bounds on
$\xi_{3}$, $\xi_4$ and $\xi_{5}$ from the perspective of a
collapsing scenario. The roots $X_{0_{1,2}}^{case3}$ have been
plotted against the parameters $\xi_{4}$ and $\xi_{5}$ in Figs.(3)
and (4).
\\\\

\begin{figure}\label{f2}
~~~~~~~~~\includegraphics[height=1.7in,width=2.5in]{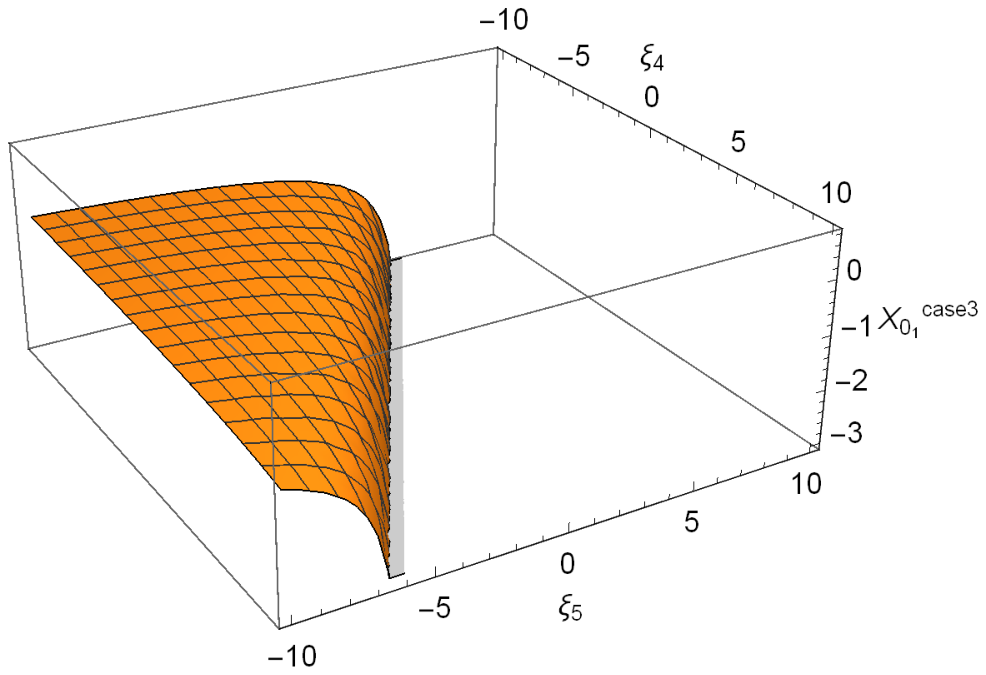}~~~~~~~~~~~~~~~~~~\includegraphics[height=1.7in,width=2.5in]{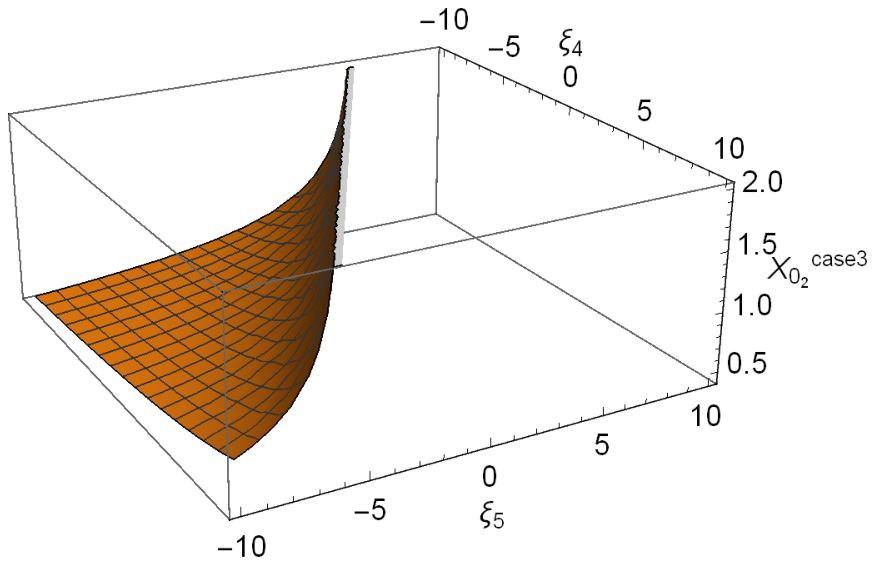}~~~~~~~\\

~~~~~~~~~~~~~~~~~~~~~~~~~~~~Fig.3~~~~~~~~~~~~~~~~~~~~~~~~~~~~~~~~~~~~~~~~~~~~~~~~~~~~~~~~~~~~~Fig.4~~~~~~~~~\\

\vspace{1mm} \textit{\textbf{Figs.3 and 4} show the variation of
the collapse parameter $X_{0}$ for different values of $\xi_4$ and
$\xi_5$ for Case-3 of Model-1. Fig.3 shows the variation for the
first root $X_{0_1}^{case3}$, whereas Fig.4 shows the variation
for the second root $X_{0_2}^{case3}$. Here we have taken
$\xi_{3}=0.5$.}
\end{figure}

\subsubsection{Case-4}
Using Eq.(\ref{solution5}) in Eq.(\ref{X0}) we get,
\begin{eqnarray}\label{X04}
\frac{2}{X_{0}}=
\begin{array}llim\\
\begin{tiny}t\rightarrow 0\end{tiny}\\
\begin{tiny}r\rightarrow 0\end{tiny}
\end{array}
\left[1+\frac{g_{1}}{2nr\left(\omega+3g_{2}\omega-g_{2}\right)}\right]
\end{eqnarray}

Just like Case-2, here also we get $\frac{2}{X_{0}}\rightarrow
\infty$, which implies $X_{0}\rightarrow 0$. The mass function
being independent of $t$ and $r$ does not generate a realistic
collapsing scenario for this particular case according to our
scheme of study.

\subsection{Model-2}

\subsubsection{Case-1}
In this model we will replace $X_{0}$ by $Y_{0}$, just to
differentiate the results from those obtained for model-1.
Moreover this is just a representational issue. The definition
remains same as given in Eq.(\ref{X0}). Using Eq.(\ref{solution6})
in Eq.(\ref{X0}) we get,
\begin{eqnarray}\label{X05}
\frac{2}{Y_{0}}=
\begin{array}llim\\
\begin{tiny}t\rightarrow 0\end{tiny}\\
\begin{tiny}r\rightarrow 0\end{tiny}
\end{array}
\left[1-\frac{h_{6}(t)}{r}AiryAi\left[-\frac{2^{1/3}nr}{g_{1}\left(-n/g_{1}\right)^{2/3}}\right]
-\frac{h_{7}(t)}{r}AiryBi\left[-\frac{2^{1/3}nr}{g_{1}\left(-n/g_{1}\right)^{2/3}}\right]\right]
\end{eqnarray}
Here we consider: ~~~$h_{6}(t)=\xi_{6}t$,
~~$h_{7}(t)=\xi_{7}t$,~~~ where $\xi_{6}$ and $\xi_{7}$ arbitrary
constants. Using these functional forms in Eq.(\ref{X05}) we get
the following algebraic equation for this case,
\begin{equation}\label{algebraic3}
\frac{1}{\Gamma(2/3)}\left(\frac{\xi_{6}}{3^{2/3}}+\frac{\xi_{7}}{3^{1/6}}\right)Y_{0}^{2}-Y_{0}+2=0
\end{equation}
It should be noted that this equation is similar to the one
obtained for Case-1 in model-1. This is due to the fact that,
though the mass functions have different forms in the two cases,
yet their limiting values coincide with other and hence generate
similar collapsing scenario. The solution for the above equation
is obtained as,
\begin{equation}\label{X0sol3}
Y_{0_{1,2}}^{case1}=\frac{\sqrt{\Gamma(2/3)}}{2\sqrt{3}\left(\frac{\xi_{6}}{3^{2/3}}+\frac{\xi_{7}}{3^{1/6}}
\right)}\left[\sqrt{3\Gamma(2/3)}\pm\sqrt{3\Gamma(2/3)-8\times
3^{1/3}\xi_{6}-8\times 3^{5/6}\xi_{7}}\right]
\end{equation}
where $3\Gamma(2/3)\geq 8\times 3^{1/3}\xi_{6}+8\times
3^{5/6}\xi_{7}$.
The collapsing outcomes may be discussed as below,\\

Conditions for a local NS:~~~~~  $Y_{0_{1}}^{case1}>0$~~ $\&$~~  $Y_{0_{2}}^{case1}<0$~~~~~OR~~~~~$Y_{0_{1}}^{case1}<0$~~ $\&$~~  $Y_{0_{2}}^{case1}>0$\\

Conditions for global NS:~~~~~~   $Y_{0_{1}}^{case1}>0$ ~~ $\&$~~ $Y_{0_{2}}^{case1}>0$\\

Condition for BH:~~~~~~~~~~~~~~~~   $Y_{0_{1}}^{case1}<0$ ~~$\&$~~$Y_{0_{2}}^{case1}<0$\\\\
The above conditions put numerical bounds on $\xi_6$ and $\xi_{7}$
from the perspective of a collapsing scenario. The roots
$Y_{0_{1,2}}^{case1}$ have been plotted against the parameters
$\xi_{6}$ and $\xi_{7}$ in Figs.(5)
and (6).\\\\

\begin{figure}\label{f3}
~~~~~~~~~\includegraphics[height=1.7in,width=2.5in]{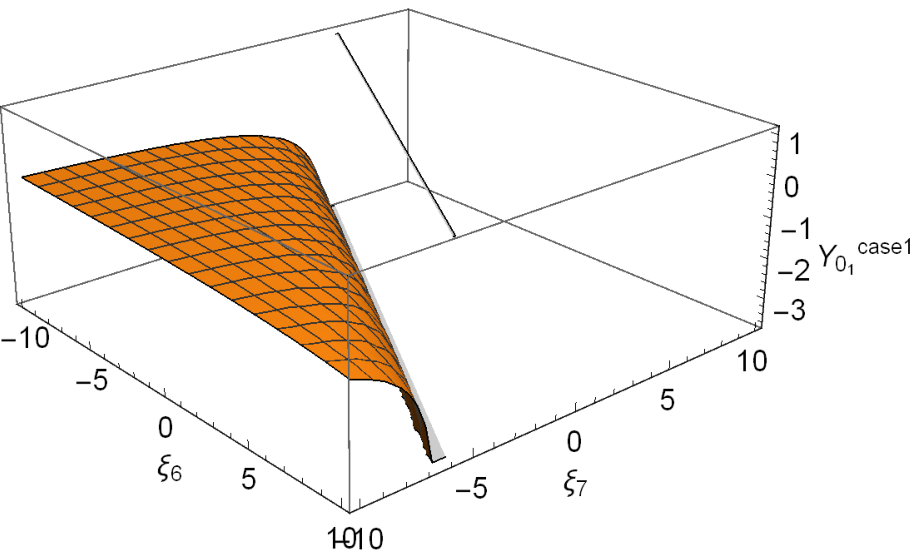}~~~~~~~~~~~~~~~~~~\includegraphics[height=1.7in,width=2.5in]{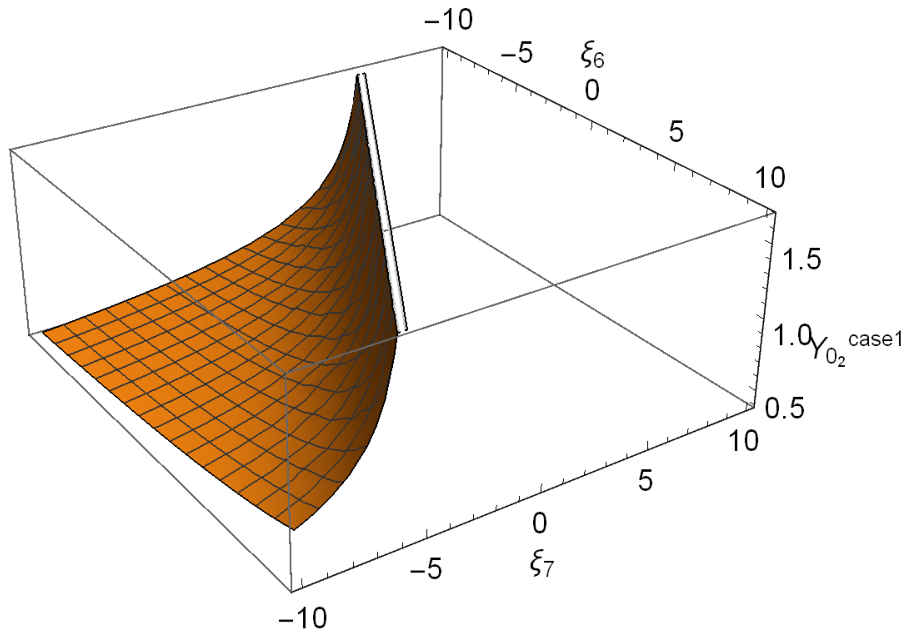}~~~~~~~\\

~~~~~~~~~~~~~~~~~~~~~~~~~~~~Fig.5~~~~~~~~~~~~~~~~~~~~~~~~~~~~~~~~~~~~~~~~~~~~~~~~~~~~~~~~~~~~~Fig.6~~~~~~~~~\\

\vspace{1mm} \textit{\textbf{Figs.5 and 6} show the variation of
the collapse parameter $Y_{0}$ for different values of $\xi_6$ and
$\xi_7$ for Case-1 of Model-2. Fig.5 shows the variation for the
first root $Y_{0_1}^{case1}$, whereas Fig.6 shows the variation
for the second root $Y_{0_2}^{case1}$.}
\end{figure}

\subsubsection{Case-2}
Using Eq.(\ref{solution7}) in Eq.(\ref{X0}) we get,
\begin{eqnarray}\label{X06}
\frac{2}{Y_{0}}=
\begin{array}llim\\
\begin{tiny}t\rightarrow 0\end{tiny}\\
\begin{tiny}r\rightarrow 0\end{tiny}
\end{array}
\left[1+\frac{g_{1}}{2nr}\right]
\end{eqnarray}
Here we get $\frac{2}{Y_{0}}\rightarrow \infty$, which implies
$Y_{0}\rightarrow 0$. The mass function being independent of $t$
and $r$ does not generate a realistic collapsing scenario for this
particular case according to our scheme of study. This is
equivalent to the scenarios in Case-2 and Case-4 in model-1.

\subsubsection{Case-3}
Using Eq.(\ref{solution8}) in Eq.(\ref{X0}) we get,
\begin{eqnarray}\label{X07}
\frac{2}{Y_{0}}=
\begin{array}llim\\
\begin{tiny}t\rightarrow 0\end{tiny}\\
\begin{tiny}r\rightarrow 0\end{tiny}
\end{array}
\left[1-\frac{h_{8}(t)}{r}-h_{9}(t)\right]
\end{eqnarray}
Here we consider  $h_{8}(t)=\xi_{8}t$ (where $\xi_{8}$ is an
arbitrary constant). We do not need to consider any particular
functional form for $h_{9}(t)$. This is because irrespective of
the form of $h_{9}(t)$, it will always yield a constant value in
the limit $t\rightarrow 0$. This gives an additional degree of
freedom to the collapsing system. The above equation yields,
\begin{equation}\label{algebraic4}
\xi_{8}Y_{0}^2+\left(\xi_{9}-1\right)Y_{0}+2=0
\end{equation}
where $\xi_{9}$ is the limiting value of $h_{9}(t)$ as
$t\rightarrow 0$. The above equations yields the solution,
\begin{equation}\label{X0sol4}
Y_{0_{1,2}}^{case3}=\frac{1-\xi_{9}\pm
\sqrt{\left(\xi_{9}-1\right)^{2}-8\xi_{8}}}{2\xi_{8}}
\end{equation}
where $\left(\xi_{9}-1\right)^{2}\geq 8\xi_{8}$. The conditions
for different collapse outcomes are given below,\\

Conditions for a local NS:~~~~~  $Y_{0_{1}}^{case3}>0$~~ $\&$~~  $Y_{0_{2}}^{case3}<0$~~~~~OR~~~~~$Y_{0_{1}}^{case3}<0$ ~~ $\&$~~ $Y_{0_{2}}^{case3}>0$\\

Conditions for global NS:~~~~~~   $Y_{0_{1}}^{case3}>0$ ~~ $\&$~~ $Y_{0_{2}}^{case3}>0$\\

Condition for BH:~~~~~~~~~~~~~~~~   $Y_{0_{1}}^{case3}<0$ ~~$\&$~~$Y_{0_{2}}^{case3}<0$\\\\
The above conditions put numerical bounds on $\xi_8$ and $\xi_{9}$
from the perspective of gravitational collapse of a massive star.
The roots $Y_{0_{1,2}}^{case3}$ have been plotted against the
parameters $\xi_{8}$ and $\xi_{9}$ in Figs.(7)
and (8).\\

\begin{figure}\label{f4}
~~~~~~~~~\includegraphics[height=1.7in,width=2.5in]{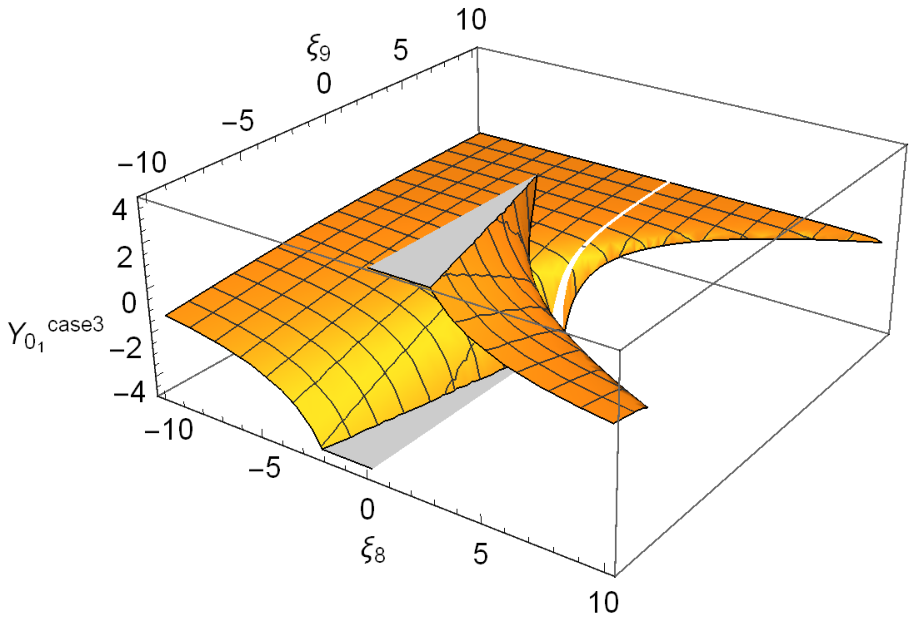}~~~~~~~~~~~~~~~~~~\includegraphics[height=1.7in,width=2.5in]{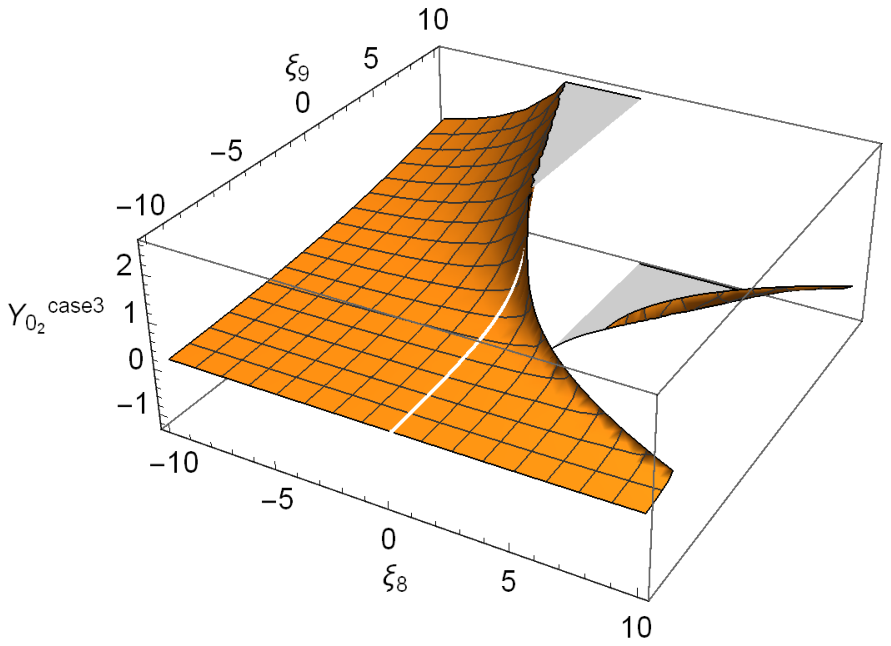}~~~~~~~\\

~~~~~~~~~~~~~~~~~~~~~~~~~~~~Fig.7~~~~~~~~~~~~~~~~~~~~~~~~~~~~~~~~~~~~~~~~~~~~~~~~~~~~~~~~~~~~~Fig.8~~~~~~~~~\\

\vspace{1mm} \textit{\textbf{Figs.7 and 8} show the variation of
the collapse parameter $Y_{0}$ for different values of $\xi_8$ and
$\xi_9$ for Case-3 of Model-2. Fig.7 shows the variation for the
first root $Y_{0_1}^{case3}$, whereas Fig.8 shows the variation
for the second root $Y_{0_2}^{case3}$.}
\end{figure}

\subsubsection{Case-4}
Using Eq.(\ref{solution9}) in Eq.(\ref{X0}) we get,
\begin{eqnarray}\label{X08}
\frac{2}{Y_{0}}=
\begin{array}llim\\
\begin{tiny}t\rightarrow 0\end{tiny}\\
\begin{tiny}r\rightarrow 0\end{tiny}
\end{array}
\left[1-\frac{r^{2}}{6}-\frac{h_{10}(t)}{r}\right]
\end{eqnarray}
Here we consider the functional form~ $h_{10}(t)=\xi_{10}t$, where
$\xi_{10}$ is an arbitrary constant. From the above equation we
get the algebraic equation,
\begin{equation}\label{algebraic5}
\xi_{10}Y_{0}^{2}-Y_{0}+2=0
\end{equation}
Solving the above equation we get,
\begin{equation}\label{X0sol5}
Y_{0_{1,2}}^{case4}=\frac{1\pm \sqrt{1-8\xi_{10}}}{2\xi_{10}}
\end{equation}
where $\xi_{10}\leq 1/8$. Here the conditions for NS and BH can be
discussed as below,\\

Conditions for a local NS:~~~~~  $Y_{0_{1}}^{case4}>0$~~ $\&$~~  $Y_{0_{2}}^{case4}<0$~~~~~OR~~~~~$Y_{0_{1}}^{case4}<0$~~ $\&$~~  $Y_{0_{2}}^{case4}>0$\\

Conditions for global NS:~~~~~~   $Y_{0_{1}}^{case4}>0$ ~~ $\&$~~ $Y_{0_{2}}^{case4}>0$\\

Condition for BH:~~~~~~~~~~~~~~~~   $Y_{0_{1}}^{case4}<0$ ~~$\&$~~$Y_{0_{2}}^{case4}<0$\\\\
The above conditions put numerical bounds on $\xi_{10}$ from the
perspective of gravitational collapse of a massive star. The roots
$Y_{0_{1,2}}^{case4}$ have been plotted against the parameter
$\xi_{10}$ in Fig.(9).\\

\begin{figure}\label{f5}
~~~~~~~~~~~~~~~~~~~~~~~~~~~~~~~~~~\includegraphics[height=1.7in,width=2.5in]{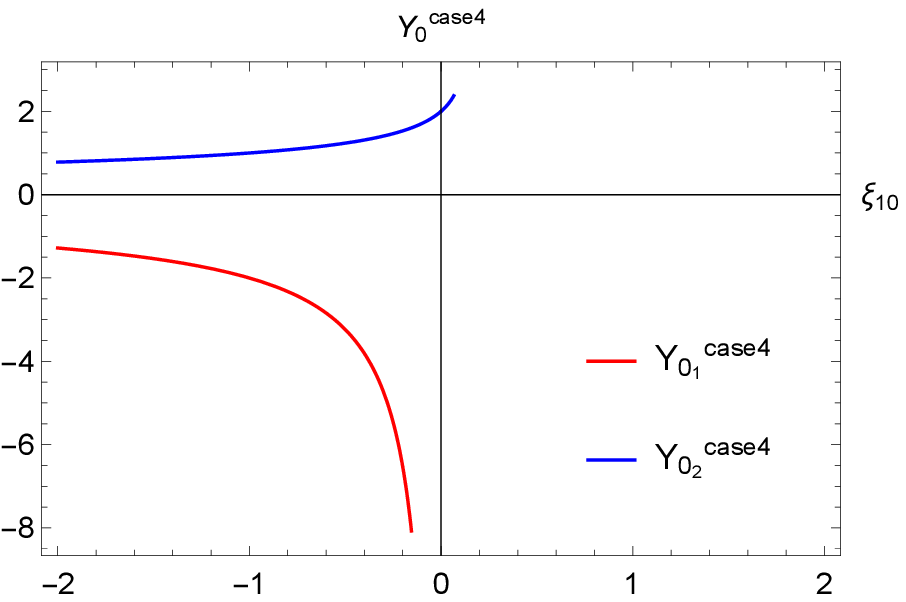}~~~~~~~~~~~\\

~~~~~~~~~~~~~~~~~~~~~~~~~~~~~~~~~~~~~~~~~~~~~~~~~~~Fig.9~~~~~~~~~~~~~~~~~~~~~~~~\\

\vspace{1mm} \textit{\textbf{Fig.9} shows the variation of the
collapse parameter $Y_{0}$ for different values of $\xi_{10}$ for
Case-4 of Model-2.}
\end{figure}

\subsection{Numerical Analysis}
In order get greater insights about the nature of the singularity
formed as an end state of the gravitational collapse for our
models, we have generated plots for the collapsing parameter
($X_{0}$ or $Y_{0}$) against the other free parameters. According
to our scheme of study, we are interested only in the signature of
the collapse parameter ($X_{0}$ or $Y_{0}$) and not in the actual
value of the parameter. In figs.(1) and (2) we have generated
plots for the Case-1 (DP model) of Model-1. We see from the
figures that the first root lies in the negative region, whereas
the second root lies in the positive region. So it can predicted
that for this case, we will have a local NS. In figs.(3) and (4)
plots have been obtained for Case-3 (PE model) of Model-1. We see
that for the considered initial conditions the first root is again
negative and the second root is positive. So here also the
collapse will end in a local NS. Figures (5) and (6) show the
corresponding plots for Case-1 (TP model) of Model-2. In fig.(5)
although the major portion of the surface lies in the negative
region, yet there is an array of points represented by a straight
line lying in the positive region around $\xi_{7}=0$. In fig.(6)
the entire surface lies in the positive region. So in this case,
there is a possibility to get more than a local NS. By properly
adjusting the initial conditions, it is quite possible that more
than one null geodesic originating in the singularity reach a
distant observer. In case of such an event, the NS will become
global in nature. In figs.(7) and (8) we have obtained plots for
Case-3 (DPE model) of Model-2. Here we see that by properly
adjusting the parameters $\xi_{8}$ and $\xi_{9}$ both positive and
negative values can be realized for both the roots. So in this
case we can have BH, local NS and global NS depending on the
chosen initial conditions. Finally in fig.(9) we have obtained
plots for Case-4 (TE model) of Model-2. Here we see that
irrespective of the initial conditions one root is always positive
and the other is always negative. So here the singularity is
destined to be a local NS.

\section{Strength of the singularity (Curvature growth near the singularity)}
The gravitational strength of a singularity is defined as the
estimate of its destructive capacity. We know that most theories
of gravity till date have been plagued by the existence of
singularities. Though theoretical methods of removal of such
singularities have been proposed in literature, yet they are
highly exotic in nature and far from being comprehensible. It is
known that singularities are holes in the fabric of the otherwise
continuous and smooth spacetime. Now for a weak singularity the
hole is shallow and an extension of space-time is possible through
the singularity. This is equivalent to a removable discontinuity
mathematically and can be a cure for the discontinuity of
spacetime at a singularity. From the above discussion it is quite
clear that one should be highly interested in finding out whether
a singularity is strong or weak in nature. According to Tipler
\cite{Tipler} a curvature singularity is said to be strong if any
object hitting it is squeezed to zero volume. In Ref.\cite{Tipler}
the condition for a strong singularity has been given as,
\begin{eqnarray}
\begin{array}{c}
S=\lim \tau^{2}\psi \\
\begin{tiny}\tau\rightarrow 0\end{tiny}\\
\end{array}
\begin{array}{c}
=\lim \tau^{2}R_{\mu\nu}K^{\mu}K^{\nu}>0 \\
\begin{tiny}\tau\rightarrow 0\end{tiny}\\
\end{array}
\end{eqnarray}
where $R_{\mu\nu}$ is the Ricci tensor, $\psi$ is a scalar given
by the relation $\psi=R_{\mu\nu}K^{\mu}K^{\nu}$, where
$K^{\mu}=dx^{\mu}/d\tau$ represents the tangent to the non
spacelike geodesics at the singularity and $\tau$ is the affine
parameter. In Ref.\cite{Maharaj} Mkenyeleye et al. have shown
that,
\begin{eqnarray}\label{maha}
\begin{array}{c}
S=\lim \tau^{2}\psi \\
\begin{tiny}\tau\rightarrow 0\end{tiny}\\
\end{array}
\begin{array}{c}\label{stren}
=\frac{1}{4}X_{0}^{2}\left(2\dot{m_{0}}\right) \\
\begin{tiny}~\end{tiny}\\
\end{array}
\end{eqnarray}
where
\begin{eqnarray}
\begin{array}{c}
m_{0}=\lim~ m(t,r) \\
\begin{tiny}t\rightarrow 0\end{tiny}\\
\begin{tiny}r\rightarrow 0\end{tiny}
\end{array}
\end{eqnarray}
and
\begin{eqnarray}\label{massd}
\begin{array}{c}
\dot{m_{0}}=\lim \frac{\partial}{\partial~t}\left(m(t,r)\right) \\
\begin{tiny}t\rightarrow 0\end{tiny}\\
\begin{tiny}r\rightarrow 0\end{tiny}
\end{array}
\end{eqnarray}
In ref. \cite{Maharaj} it has also been shown that the relation
between $X_{0}$ and the limiting values of mass is given by,
\begin{equation}\label{xmass}
X_{0}=\frac{2}{1-2m_{0}'-2\dot{m_{0}}X_{0}}
\end{equation}
where
\begin{eqnarray}\label{dashedmass}
\begin{array}{c}
m_{0}'=\lim \frac{\partial}{\partial~r}\left(m(t,r)\right) \\
\begin{tiny}t\rightarrow 0\end{tiny}\\
\begin{tiny}r\rightarrow 0\end{tiny}
\end{array}
\end{eqnarray}
and $\dot{m_{0}}$ is given by the eqn.(\ref{massd}).

Studies by Dwivedi and Joshi in Refs.\cite{strongcurvature, cch2}
showed that any classical singularity in Vaidya spacetime in
Einstein gravity is supposed to be a strong curvature singularity
in a very strong sense. Additionally they have also shown that the
conjecture \cite{tiplernew} that the strong curvature
singularities are never naked is not always true. It is speculated
that in the background of $f(R,T)$ gravity the strength of the
singularity may weaken due to the exotic component arising from
the modified gravity. Moreover the structure of such a NS was
studied in detail in Ref.\cite{strongcurvature1} and it was shown
that the singularity admits a directional behaviour in terms of
curvature growth along the geodesics terminating in the
singularity. On the contrary it was found that in a quantum regime
the singularity formed is supposed to be gravitationally weak,
thus allowing a continuous extension of the spacetime beyond the
singularity \cite{weakness}. Below we study the strength of the
singularities for the different models.

\subsection{Model-1}

\subsubsection{Case-1}
Using Eqs.(\ref{solution1}), (\ref{stren}) and (\ref{massd}) we
have,
\begin{equation}\label{strength1}
S=\lim_{\tau\rightarrow 0}
\tau^{2}\psi=\frac{X_{0}^{2}}{2\Gamma(2/3)}\left(\frac{\xi_{1}}{3^{2/3}}+\frac{\xi_{2}}{3^{1/6}}\right)
\end{equation}
It is obvious that the signature of the above expression is
independent of the collapsing parameter $X_{0}$ since
$X_{0}^{2}>0$. So the strength of the singularity ultimately
depends on the values of the parameters $\xi_{1}$ and $\xi_{2}$.
The condition for a strong singularity is
$\frac{1}{\Gamma(2/3)}\left(\frac{\xi_{1}}{3^{2/3}}+\frac{\xi_{2}}{3^{1/6}}\right)>0$
and that for a weak singularity is
$\frac{1}{\Gamma(2/3)}\left(\frac{\xi_{1}}{3^{2/3}}+\frac{\xi_{2}}{3^{1/6}}\right)\leq
0$.

\subsubsection{Case-2}
Using Eqs.(\ref{solution3}), (\ref{stren}) and (\ref{massd}) we
have,
\begin{equation}\label{strength2}
S=\lim_{\tau\rightarrow 0} \tau^{2}\psi=0
\end{equation}
The above value shows that the singularity formed is weak in
nature. In the previous section it was seen that for this model,
the constancy of mass parameter did not assist in studying the
nature of the singularity. But whatever be the nature of the
singularity formed, it should always be gravitationally weak in
nature.

\subsubsection{Case-3}
Using Eqs.(\ref{solution4}), (\ref{stren}) and (\ref{massd}) we
have,
\begin{equation}\label{strength3}
S=\lim_{\tau\rightarrow 0}
\tau^{2}\psi=\frac{X_{0}^{2}}{2\Gamma(2/3)}\left(\frac{\xi_{4}}{3^{2/3}}+\frac{\xi_{5}}{3^{1/6}}\right)
\end{equation}
Similar to case-1, here the strength of singularity is independent
of $X_{0}$. The condition for a strong singularity is
$\frac{1}{\Gamma(2/3)}\left(\frac{\xi_{4}}{3^{2/3}}+\frac{\xi_{5}}{3^{1/6}}\right)>0$
and that for a weak singularity is
$\frac{1}{\Gamma(2/3)}\left(\frac{\xi_{4}}{3^{2/3}}+\frac{\xi_{5}}{3^{1/6}}\right)\leq
0$.

\subsubsection{Case-4}
Using Eqs.(\ref{solution5}), (\ref{stren}) and (\ref{massd}) we
have,
\begin{equation}\label{strength4}
S=\lim_{\tau\rightarrow 0} \tau^{2}\psi=0
\end{equation}
This is a situation similar to case-2 where the singularity is
always gravitationally weak.

\subsection{Model-2}

\subsubsection{Case-1}
Using Eqs.(\ref{solution6}), (\ref{stren}) and (\ref{massd}) we
have,
\begin{equation}\label{strength5}
S=\lim_{\tau\rightarrow 0}
\tau^{2}\psi=\frac{Y_{0}^{2}}{2\Gamma(2/3)}\left(\frac{\xi_{6}}{3^{2/3}}+\frac{\xi_{7}}{3^{1/6}}\right)
\end{equation}
Here the signature of the above expression and hence the strength
of the singularity depends on the values of the parameters
$\xi_{6}$ and $\xi_{7}$. We get a strong singularity if
$\frac{1}{\Gamma(2/3)}\left(\frac{\xi_{6}}{3^{2/3}}+\frac{\xi_{7}}{3^{1/6}}\right)>0$,
and a weak singularity if
$\frac{1}{\Gamma(2/3)}\left(\frac{\xi_{6}}{3^{2/3}}+\frac{\xi_{7}}{3^{1/6}}\right)\leq
0$.

\subsubsection{Case-2}
Using Eqs.(\ref{solution7}), (\ref{stren}) and (\ref{massd}) we
have,
\begin{equation}\label{strength6}
S=\lim_{\tau\rightarrow 0} \tau^{2}\psi=0
\end{equation}
Hence the singularity is gravitationally weak in nature.

\subsubsection{Case-3}
Using Eqs.(\ref{solution8}), (\ref{stren}) and (\ref{massd}) we
have,
\begin{equation}\label{strength7}
S=\lim_{\tau\rightarrow 0}
\tau^{2}\psi=\frac{1}{2}Y_{0}^{2}\xi_{8}
\end{equation}
Here we have considered no special form for the function
$h_{9}(t)$ as was done in the previous section. It is clear from
the above expression that the strength of the singularity
basically depends on the signature of $\xi_{8}$. If $\xi_{8}>0$,
then the singularity is strong and if $\xi_{8}\leq 0$, then the
singularity is weak in nature. However if we do consider a special
form for the function $h_{9}(t)$, we can have a different result.
If we consider $h_{9}(t)=\gamma_{9}\log(t)$, then we have from the
Eqs.(\ref{solution8}), (\ref{stren}) and (\ref{massd}),
\begin{equation}\label{strength8}
S=\lim_{\tau\rightarrow 0}
\tau^{2}\psi=\frac{1}{2}Y_{0}^{2}\left(\xi_{8}+\frac{\gamma_{9}}{Y_{0}}\right)
\end{equation}
Now using Eqs.(\ref{solution8}), (\ref{massd}), (\ref{xmass}) and
(\ref{dashedmass}) we get a relation from where the values of
$X_{0}$ may be extracted. Using these values of $X_{0}$ in the
above equation we may have a different scenario for the strength
of the singularity.

\subsubsection{Case-4}
Using Eqs.(\ref{solution9}), (\ref{stren}) and (\ref{massd}) we
have,
\begin{equation}\label{strength9}
S=\lim_{\tau\rightarrow 0}
\tau^{2}\psi=\frac{1}{2}Y_{0}^{2}\xi_{10}
\end{equation}
Here the strength of the singularity depends on the signature of
$\xi_{10}$. If $\xi_{10}>0$, then the singularity is strong, and
if $\xi_{10}\leq 0$, the singularity is weak.

\section{Conclusion and Discussion}
In this work, we have explored a gravitational collapse mechanism
of a massive star in $f(R,T)$ gravity. A time dependent Vaidya
spacetime is used to model the collapsing phenomenon. The
Einstien's field equations for $f(R,T)$ gravity in the Vaidya
spacetime are calculated and the corresponding solutions for the
mass parameter $m(t,r)$ are obtained. We have considered two
different category of $f(R,T)$ models, each consisting of four
sub-models. The two models are considered on the basis of the
nature of coupling between the scalar invariants $R$ and $T$. The
sub-models for each model basically involve various combinations
of power and exponential functional forms. Here we considered the
collapse of a massive star $(>20 M_{\odot})$, which will
invariably continue its collapse until the formation of a
singularity. The huge mass of the parent star will always keep the
collapsing mass beyond the Chandrasekhar limit $(1.3 M_{\odot})$,
and hence neither the electron nor the neutron degeneracy pressure
will be able to counterbalance the inward the collapsing force.
Hence the collapse will not terminate in any middle stage like a
white dwarf or a neutron star, but will continue all the way to a
singularity (BH or NS). The scheme followed for the gravitational
collapse study involved the quest for outgoing radial null
geodesics from the central singularity formed as an end state of
the collapse. If such outgoing geodesics exists then the
singularity becomes a naked singularity and the formation of the
event horizon is hindered. Such a situation will definitely defy
the cosmic censorship hypothesis. Moreover depending on the number
of such escaping geodesics, we can have a locally or globally
naked singularity. More number of escaping geodesics will mean
greater exposure time of the singularity to an external observer,
and hence result in a globally naked singularity. However if no
such geodesic escape from the singularity, the collapse is
destined to end in a black hole and thus favour the censorship
hypothesis. Our study predicts that in almost all the cases of
model-1 we get a locally naked singularity. Model-2 seems to be a
mixed bag, predicting the formation of black holes, local and
global naked singularities depending on the initial conditions.
However in the Case-4 of Model-2 (TE model), the collapse always
results in a local naked singularity. So here it should be noted
that the nature of coupling between the scalar invariants $R$ and
$T$ does play a very important role in the nature of singularity
formed as an end state of the collapse. For minimal coupling
(Model-1), we see that the collapse generally ends in a local
naked singularity. But for non-minimal coupling (Model-2), all the
options (BH, local and global NS) are possible except the TE model
(Case-4). Hence these models resulting from the minimal and
non-minimal coupling between curvature and matter, can be
considered as significant counterexamples of the cosmic censorship
hypothesis. But as we know that non-minimal coupling is
observationally the favoured model, the result derived for this
model-2 will be cosmologically more relevant. Moreover we see that
for minimal coupling we generally do not get the global nature of
the naked singularity, but in case of non-minimal coupling this
can be a reality. One thing which may be worrying for the reader
is that the final limiting forms ($X_{0}$ or $Y_{0}$) in the
collapsing scheme does not involve the model parameters $g_{i}$ or
$\beta_{i}$, $i=1,2,3$. So how does one differentiate the collapse
outcomes between the models? We see that here the solutions are in
terms of Airy functions which are relatively complicated
mathematical forms. In the limiting scenario the argument of these
functions vanish giving constant values, which is reason we do not
see any model parameters in the limiting forms. However it should
be mentioned here that the functional forms of $X_{0}$ or $Y_{0}$
are different for different models, which is testimony of the fact
they arise from different functional forms. Moreover the imprints
of such functional forms are carried by the functions $h_{i}(t)$,
$i=1,2,3...10$, and thus the parameters $\xi_{i}$, $i=1,2,3...10$
which are present in the analysis. One more thing that the reader
needs to note is that some of the solutions derived for the
special cases in model-2 are valid for early universe. So in such
cases we are actually studying the collapsing scenarios of
primordial black holes that existed at the beginning of the
universe. However it should also be kept in mind that these
solutions are just specific examples to get greater insights into
the bigger picture and in no sense represent the entire story.

To complement the collapsing scheme we have studied the strength
of the singularity formed for all our models. We see that for the
DP (case-1) and PE (case-3) models of model-1, the singularity can
be both gravitationally weak or strong depending on the model
parameters. However for the DE (case-2) and EP (case-4) models,
the singularity formed is always gravitationally weak. A weak
singularity will obviously be pathologically favoured because the
spacetime can be extended beyond such a singularity and we get a
sense of continuity. For model-2, we see that for the TP (case1),
DPE (case-3) and TE (case-4) models the strength of the
singularity depends on the initial conditions but for the DEP
(case-2) model the singularity is always gravitationally weak. So
it is understandable that for all models, by suitably adjusting
the initial conditions, we can have a sufficiently weak
singularity, which will be cosmologically desirable, since an
extension of the spacetime beyond the singularity becomes a
possibility. In principle we can create a scenario where the
singularity may be completely avoided. This is a direct
consequence of the coupling of matter with geometry and hence an
intrinsic property of $f(R,T)$ models and their exotic nature.

\section*{Acknowledgments}

The author acknowledges the Inter University Centre for Astronomy
and Astrophysics (IUCAA), Pune, India for granting visiting
associateship.

\section{Appendix}
Here we report the other components of the field equations for
this model which have not been used in our analysis.\\

\textbf{\textit{Model-1}}\\

\textbf{\textit{1. The (00)-component of field equations is given
by,}}

$$r^{4}\left\{f_{1}(R)+f_{2}(T)-2\left(f_{2}'(T)+1\right)\left(\rho+\sigma\right)+2f_{2}'(T)\omega
\rho\right\}-8f_{1}'''(R)\dot{m}'^{2}-r\left\{8f_{1}'''(R)\dot{m}'\dot{m}''\right.$$
$$\left.+r\left(f_{1}'(R)rm''+m\left(r\left(f_{1}(R)+f_{2}(T)-2\rho\left(1+f_{2}'(T)-\omega
f_{2}'(T)\right)\right)-f_{1}'(R)m''\right)-2f_{1}'(R)\dot{m}\right.\right.$$
\begin{equation}\label{fieldeq00}
\left.\left.+2f_{1}'''(R)\dot{m}''^{2}+4f_{1}''(R)\ddot{m}'+2f_{1}''(R)r\ddot{m}''\right)\right\}=0
\end{equation}

\textbf{\textit{2. The (11)-component of field equations is given
by,}}

\begin{equation}\label{fieldeq11}
f_{1}'''(R)\left[-4m'+r\left(m''+rm^{(3)}\right)\right]^2+r^{2}f_{1}''(R)\left(12m'-6rm''+r^{3}m^{(4)}\right)=0
\end{equation}
where $(3)$ and $(4)$ in the power represents the third and fourth
order derivative with respect to $r$ respectively.\\\\

\textbf{\textit{Model-2}}\\

\textbf{\textit{1. The (00)-component of field equations is given
by,}}

$$2r^{4}\sigma+2r^{3}\rho\left(r-m\right)-f_{2}(R)r^{3}\left(f_{3}(T)+2f_{3}'(T)\omega
\rho\right)\left(r-m\right)-f_{1}(R)r^{3}\left(r-m\right)+2f_{2}(R)f_{3}'(T)r^{3}
\left\{r\left(\rho+\sigma\right)-\rho m\right\}$$
$$+\left(f_{1}'(R)+f_{2}'(R)f_{3}(T)\right)r^{2}\left\{m''\left(r-m\right)-2\dot{m}\right\}
+2f_{2}'(R)f_{3}''(T)r^{4}\lambda^{2}\dot{\rho}^{2}+4f_{2}''(R)f_{3}'(T)r^{2}\lambda\dot{\rho}\left(2\dot{m}'+r\dot{m}''\right)$$

$$+2f_{1}'''(R)\left(2\dot{m}'+r\dot{m}''\right)^{2}+
2f_{2}'''(R)f_{3}(T)\left(2\dot{m}'+r\dot{m}''\right)^{2}+2f_{2}'(R)f_{3}'(T)r^{4}\lambda\ddot{\rho}
+4f_{1}''(R)r^{2}\ddot{m}'+4f_{2}''(R)f_{3}(T)r^{2}\ddot{m}'$$
\begin{equation}\label{fieldeq00m2}
+2\left(f_{1}''(R) +f_{2}''(R)f_{3}(T)\right)r^{3}\ddot{m}''=0
\end{equation}

\textbf{\textit{2. The (11)-component of field equations is given
by,}}

$$r^{6}\left(-f_{2}'(R)f_{3}''(T)\lambda^{2}(\rho')^{2}-f_{2}'(R)f_{3}'(T)\lambda\rho''\right)-r^{3}
\left[2f_{2}''(R)f_{3}'(T)\lambda\rho'\left(-4m'+r\left(m''+rm'''\right)\right)\right]$$
$$-f_{1}'''(R)\left[-4m'+r\left(m''+rm'''\right)\right]^{2}-f_{2}'''(R)f_{3}(T)
\left[-4m'+r\left(m''+rm'''\right)\right]^{2}-r^{2}f_{1}''(R)\left(12m'-6rm''+r^{3}m^{iv}\right)$$
\begin{equation}\label{fieldeq11m2}
- r^{2}f_{2}''(R)f_{3}(T)\left(12m'-6rm''+r^{3}m^{iv}\right)=0
\end{equation}

\textbf{\textit{Airy Function}}\cite{airy}\\

Here we would like to present a short description of Airy
functions for the reader's convenience. Airy function is a special
function named after the British astronomer George Biddell Airy
(1801-1892). There are in fact, two Airy functions $Ai(x)$ (Airy
function of the first kind) and $Bi(x)$ (Airy function of the
second kind), which are linearly independent solutions of the Airy
differential equation given by,
\begin{equation}\label{airyeq}
\frac{d^{2}y}{dx^{2}}-xy=0
\end{equation}
For real values of $x$ the Airy function of the first kind is
defined by the improper integral,
\begin{equation}\label{airydef1}
Ai(x)=\frac{1}{\pi}\int_{0}^{\infty}\cos\left(\frac{t^{3}}{3}+xt\right)dt\equiv
\frac{1}{\pi}\lim_{b\rightarrow
\infty}\int_{0}^{b}\cos\left(\frac{t^{3}}{3}+xt\right)dt
\end{equation}
which is convergent. This solution is subject to the condition
$y\rightarrow 0$ as $x\rightarrow \infty$. The Airy function of
the second kind is defined as,
\begin{equation}\label{airydef2}
Bi(x)=\frac{1}{\pi}\int_{0}^{\infty}\left[exp\left(-\frac{t^3}{3}+xt\right)
+\sin\left(\frac{t^3}{3}+xt\right)\right]dt
\end{equation}
This solution has the same amplitude of oscillation as Ai(x) as
$x\rightarrow -\infty$ differing in phase by $\pi/2$. The values
of Airy function $(Ai(x), Bi(x))$ and its derivatives $(Ai'(x),
Bi'(x))$ at $x=0$ are given by,
\begin{equation}\label{airy0}
Ai(0)=\frac{1}{3^{2/3}\Gamma(2/3)}, ~~~~
Bi(0)=\frac{1}{3^{1/6}\Gamma(2/3)},
~~~Ai'(0)=-\frac{1}{3^{1/3}\Gamma(1/3)},
~~~Bi'(0)=\frac{3^{1/6}}{\Gamma(1/3)}
\end{equation}



\begin{thebibliography}{99}

\bibitem{acc1} S. Perlmutter et. al. :- {\it Astrophys. J.} {\bf 517} 565 (1999).
\bibitem{acc2} A. G. Riess et al. :- {\it Astron. J.} {\bf 116} 1009 (1998).
\bibitem{mod1} S. Nojiri, S. D. Odintsov, V. K. Oikonomou :- {\it Phys. Rep.} {\bf 692} 1 (2017).
\bibitem{mod2} S. Nojiri and S. D. Odintsov :- {\it Int. J. Geom. Methods Mod. Phys.} {\bf 04} 115 (2007).
\bibitem{mod3} S. Capozziello, R. D'Agostino, O. Luongo:- {\it Int. J. Mod. Phys. D} {\bf 28} 1930016 (2019).
\bibitem{de1} P. Brax :- {\it Rep. Prog. Phys.} {\bf 81} 016902 (2018)
\bibitem{fr1} T. P. Sotiriou, V. Faraoni :- {\it Rev. Mod. Phys.} {\bf 82} 451 (2010).
\bibitem{fr2} A. De Felice, S. Tsujikawa :- {\it Living Rev. Relativity} {\bf 13} 3 (2010).
\bibitem{frlm} T. Harko, F. S. N. Lobo :- {\it Eur. Phys. J. C.} {\bf 70} 373 (2010).
\bibitem{frlm2} R. Ribeiro, J. Páramos :- {\it Phys. Rev. D} {\bf 90} 124065 (2014).
\bibitem{frlm3} R. P. L. Azevedo, J. Páramos :- {\it Phys. Rev. D} {\bf 94} 064036 (2016).
\bibitem{frlm4} B. Pourhassan, P. Rudra :- {\it Phys. Rev. D} {\bf 101} 084057 (2020).
\bibitem{harko1} T. Harko, F. S. N. Lobo, S. Nojiri, S. D. Odintsov :- {\it Phys. Rev. D.} {\bf 84} 024020 (2011).
\bibitem{frt1} M. Sharif, M. Zubair :- {\it JCAP} {\bf 03} 028 (2012).
\bibitem{frt2} E. H. Baffou, M. J. S. Houndjo, M. E. Rodrigues, A. V. Kpadonou, J. Tossa :- {\it Phys. Rev. D} {\bf 92} 8, 084043 (2015).
\bibitem{frt3} P. Rudra :- {\it Eur. Phys. J. Plus} {\bf 130}  4, 66 (2015).
\bibitem{frt4} H. Shabani, M. Farhoudi :- {\t Phys. Rev. D} {\bf 88} 044048 (2013).
\bibitem{frt5} F. G. Alvarenga, A. de la Cruz-Dombriz, M. J. S. Houndjo, M. E. Rodrigues, D. Sáez-Gómez :- {\it Phys. Rev. D} {\bf 87}  10, 103526 (2013).
\bibitem{frt6} A. Das, S. Ghosh, B. K. Guha, S. Das, F. Rahaman :- {\it Phys. Rev. D} {\bf 95}  12, 124011 (2017).
\bibitem{frt7} R. Zaregonbadi, M. Farhoudi, N. Riazi :- {\it Phys. Rev. D} {\bf 94} 084052 (2016).
\bibitem{frt8} M. Sharif, A. Siddiqa :- {\it Gen. Rel. Grav.} {\bf 51}  6, 74 (2019).
\bibitem{frt9} S. Hansraj, A. Banerjee :- {\it Phys. Rev. D} {\bf 97}  10, 104020 (2018).

\bibitem{oppen} J. R. Oppenhiemer, H. Snyder :- {\it Phys. Rev.} {\bf 56} 455 (1939).
\bibitem{tolman} R. C. Tolman :- {\it Proc. Natl. Acad. Sci. USA} {\bf 20} 169 (1934);
\bibitem{bondi} H. Bondi :- {\it Mon. Not. Astron. Soc.} {\bf 107} 410 (1947).
\bibitem{collrev1} P. S. Joshi, D. Malafarina :- {\it Int. J. Mod. Phys. D} {\bf 20} 14, 2641 (2011).
\bibitem{collrev2} D. Malafarina :- {\it Universe} {\bf 3} 48 (2017).
\bibitem{penrose} R. Penrose :- {\it Riv. Nuovo Cimento} {\bf 1} 252 (1969).
\bibitem{paradox1} P. Chen, Y. C. Ong, D-h. Yeom :- {\it Phys. Rept.} {\bf 603} 1 (2015).
\bibitem{paradox2} B. Zhang, Q-y. Cai, M-s. Zhan, L. You :- {\it Phys. Rev. D} {\bf 87}  4, 044006 (2013).
\bibitem{paradox3} L. Smolin :- {\it Phys. Rev. D} {\bf 90}  2, 024074 (2014).
\bibitem{paradox4} H. Nikolic :- {\it Phys. Lett. B} {\bf 678} 218 (2009).
\bibitem{ns1} D. M. Eardley, L. Smarr :- {\it Phys. Rev. D} {\bf 19} 2239 (1979).
\bibitem{ns2} D. Christodoulou :- {\it Commun. Math. Phys.} {\bf 93} 171 (1984).
\bibitem{ns3} R. P. A. C. Newman :- {\it Class. Quantum Grav.} {\bf 3} 527 (1986).
\bibitem{ns4} I. H. Dwivedi, P. S. Joshi :- {\it Class. Quantum Grav.} {\bf 9} L39 (1992).
\bibitem{ns5} P. S. Joshi, I. H. Dwivedi :- {\it Phys. Rev. D} {\bf 47} 5357 (1993).
\bibitem{cch2} P. S. Joshi, I.H. Dwivedi :- {\it Commun. Math. Phys.} {\bf 146} 333 (1992).
\bibitem{cch3} P. S. Joshi, T.P. Singh :- {\it Phys. Rev. D} {\bf 51} 6778 (1995).
\bibitem{ns6} B. Waugh, K. Lake :- {\it Phys. Rev. D} {\bf 34} 2978 (1986).
\bibitem{ns7} A. Ori, T. Piran :- {\it Phys. Rev. D} {\bf 42} 1068 (1990).
\bibitem{lake2} K. Lake :- {\it Phys. Rev. Lett.} {\bf 68} 3129 (1992).
\bibitem{szek} P. Szekeres, V. Iyer :- {\it Phys. Rev. D} {\bf 47} 4362 (1993).
\bibitem{ghosh1} S. G. Ghosh, S. D. Maharaj :- {\it Phys. Rev. D} {\bf 85} 124064 (2012).
\bibitem{v1} P. C. Vaidya :- {\it Proc. Indian Acad. Sci. Sect. A} {\bf 33} 264 (1951).
\bibitem{s1} P. Rudra, M. Faizal, A. F. Ali :- {\it Nucl. Phys. B} {\bf 909} 725 (2016).
\bibitem{s2} Y. Heydarzade, P. Rudra, F. Darabi, A. F. Ali, M. Faizal :- {\it Phys. Lett. B} {\bf 774} 46 (2017).
\bibitem{s3} P. Rudra, S. Maity :- {\it Eur. Phys. J. C} {\bf 78} 828 (2018).
\bibitem{s4} Y. Heydarzade, P. Rudra, B. Pourhassan, M. Faizal, A. F. Ali :- {\it J. Cosmol. Astropart. Phys.} {\bf 06} 038 (2018).
\bibitem{s5} P. Rudra, U. Debnath :- {\it Can. J. Phys.} {\bf 92(11)} 1474 (2014).
\bibitem{s6} P. Rudra, R. Biswas, U. Debnath :- {\it Astrophys. Space Sci.} {\bf 354} 2 (2014).
\bibitem{s7} U. Debnath, P. Rudra, R. Biswas :- {\it Astrophys. Space Sci.} {\bf 339} 135 (2012).
\bibitem{s8} P. Rudra, R. Biswas, U. Debnath :- {\it Astrophys.Space Sci.} {\bf 335} 505 (2011).
\bibitem{s9} P. Rudra :- {\it Nucl. Phys. B.} {\bf 956} 115014 (2020).
\bibitem{land1} L. D. Landau, E. M. Lifshitz :- The Classical Theory of Fields (Butterworth-Heinemann, Oxford, 1998).
\bibitem{boer1} J. de Boer, J. Hartong, N. A. Obers, W. Sybesma, S. Vandoren :- {\it SciPost Phys.} {\bf 5} 003 (2018)
\bibitem{sing} T. P. Singh, P.S. Joshi :- {\it Class. Quant. Gravity} {\bf 13} 559 (1996).
\bibitem{Tipler} F. J. Tipler :- {\it Phys. Lett. A.} {\bf 64}, 8 (1977)
\bibitem{Maharaj} M. D. Mkenyeleye, R. Goswami, and S. D. Maharaj :- {\it Phys. Rev. D} {\bf 90}  064034 (2014).
\bibitem{strongcurvature} I. H. Dwivedi, P. S. Joshi :- {\it Class. Quantum Grav.} {\bf 6} 1599 (1989).
\bibitem{tiplernew} F. J. Tipler, C. J. S. Clarke, G. F. R. Ellis  :- {\it General Relativity and Gravitation} {\bf vol 2} p97 ed A Held (NewYork Plenum) (1980).
\bibitem{strongcurvature1} I. H. Dwivedi, P. S. Joshi :- {\it Class. Quantum Grav.} {\bf 8} 1339 (1991).
\bibitem{weakness} A. Bonanno, B. Koch, A. Platania :- {\it Foundations of Phys.} {\bf 48} 1393 (2018).
\bibitem{airy} $https://en.wikipedia.org/wiki/Airy_function$



\end{thebibliography}
\end{document}